\documentclass[12pt]{article}
\usepackage{amsfonts}
\usepackage{amssymb}
\usepackage{amsbsy}
\usepackage{mathrsfs}
\usepackage{amsmath}
\usepackage{graphicx}    % il parait que `graphicx' est requis pour `rotating'
\usepackage{rotating}    % necessaire pour avoir des gros tableaux tournes
\usepackage{epsfig}
\usepackage{color}
\input epsf.sty

\newlength{\TZ}
\newcommand{\BEQ}{\begin{equation}}     % Gleichungen Anfang ..
\newcommand{\BEA}{\begin{eqnarray}}
\newcommand{\BD}{\begin{displaymath}}
\newcommand{\EEQ}{\end{equation}}       % .. und Ende
\newcommand{\EEA}{\end{eqnarray}}
\newcommand{\ED}{\end{displaymath}}
\newcommand{\bb}{\begin{eqnarray}}
\newcommand{\ee}{\end{eqnarray}}
\newcommand{\nn}{\nonumber}

\newcommand{\vph}{\varphi}              % rundes phi
\newcommand{\vro}{\varrho}              % Variante von rho
\newcommand{\D}{{\rm d}}                % gerades d fuer Ableitungen
\newcommand{\II}{{\rm i}}               % gerades i fuer komplexe Einheit
\newcommand{\demi}{\frac{1}{2}}         % Bruch 1/2
\newcommand{\frdrei}{\frac{3d}{2}}
\newcommand{\freins}{\frac{d}{2}}
\newcommand{\wht}[1]{\widehat{#1}}      % weiter Hut
\newcommand{\lap}[1]{\overline{#1}}     % Querstrich oben
\renewcommand{\vec}[1]{\boldsymbol{#1}} % Vektoren fettgedruckt

\newcommand{\vekz}[2]
     {\mbox{${\begin{array}{c} #1  \\ #2 \end{array}}$}}
\newcommand{\fn}{\footnotesize}         % taille petite (note en bas de page)

\newcommand{\appsection}[2]{\setcounter{equation}{0}\setcounter{subsection}{0}
\section*{Appendix #1. #2}
\renewcommand{\theequation}{#1.\arabic{equation}}
              \renewcommand{\thesection}{#1} }
\catcode`\@=11
\def\numberbysection{\@addtoreset{equation}{section}
        \def\theequation{\thesection.\arabic{equation}}}
\numberbysection

\begin{document}

\begin{titlepage}

\vskip 1.5 cm
\begin{center}
{\Large \bf Exactly solvable models \\[0.15truecm] of growing interfaces and lattice gases: \\[0.16truecm]
the Arcetri models, ageing and logarithmic sub-ageing}
\end{center}

\vskip 2.0 cm
\centerline{{\bf Xavier Durang}$^{a,b}$\footnote{e-mail: xdurang1@uos.ac.kr}
and {\bf Malte Henkel}$^{c,d,e}$\footnote{e-mail: malte.henkel@univ-lorraine.fr}
}
\vskip 0.5 cm
\begin{center}
$^a$School of Physics, Korea Institute for Advanced Study, Seoul 130-722, Korea\\
$^b$Department of Physics, University of Seoul, Seoul 02504, Korea\\
$^c$Rechnergest\"utzte Physik der Werkstoffe, Institut f\"ur Baustoffe (IfB), \\ ETH Z\"urich, Stefano-Franscini-Platz 3,
CH -- 8093 Z\"urich, Switzerland\\
$^d$Groupe de Physique Statistique,
D\'epartement de Physique de la Mati\`ere et des Mat\'eriaux,
Institut Jean Lamour (CNRS UMR 7198), Universit\'e de Lorraine Nancy,
B.P. 70239, \\ F -- 54506 Vand{\oe}uvre l\`es Nancy Cedex, France\footnote{permanent address; \\
after 1$^{\rm st}$ of Januar 2018: Laboratoire de Physique et Chimie Th\'eoriques (CNRS UMR), Universit\'e de Lorraine Nancy,
B.P. 70239, F -- 54506 Vand{\oe}uvre-l\`es Nancy Cedex, France}\\
$^e$Centro de F\'{i}sica Te\'{o}rica e Computacional, Universidade de Lisboa, \\P--1749-016 Lisboa, Portugal\\~\\
\end{center}

\begin{abstract}
Motivated by an analogy with the spherical model of a ferromagnet, the three Arcetri models are defined.
They present new universality classes, either for the growth of interfaces, or else for lattice gases.
They are distinct from the common Edwards-Wilkinson and Kardar-Parisi-Zhang universality classes.
Their non-equilibrium evolution can be studied from the exact computation of their two-time correlators and responses.
The first model, in both interpretations, has a critical point in any dimension and shows simple ageing at and below criticality.
The exact universal exponents are found. The second and third model are solved at zero temperature, in one dimension,
where both show logarithmic sub-ageing, of which several distinct types are identified.
Physically, the second model describes a lattice gas and the third model interface growth.
A clear physical picture on the subsequent time- and length-scales of the sub-ageing process emerges.

~\\
%\centerline{\textcolor{red}{\Large \today}}
\end{abstract}

\vfill
%\noindent

\end{titlepage}

\setcounter{footnote}{0}

%%%%%%%%%%%%%%%%%%%%%%%%%%%%%%%%%%%%%%%%%%%%%%%%%%%%%%%%%%%%%%%%%%%%%%%%%%%%%%%%
\section{Introduction}
%%%%%%%%%%%%%%%%%%%%%%%%%%%%%%%%%%%%%%%%%%%%%%%%%%%%%%%%%%%%%%%%%%%%%%%%%%%%%%%%

The physics of the growth of interfaces is a paradigmatic example of
the emergence of non-equilibrium cooperative phenomena, with widespread applications in domains as different as
deposition of atoms on a surface, solidification, flame propagation, population dynamics, crack propagation,
chemical reaction fronts or the growth of cell colonies \cite{Bara95,Halp95,Krug97,Krie10,Corw12,Wio13,Taeu14,Wio17}.
Several universal growth and roughness exponents characterise the morphology of the growing interface and the time-dependent
properties are quite analogous to phenomena encountered in the
physical ageing in glassy and non-glassy systems \cite{Cugl03,Henk10,Taeu17}.
Several universality classes of interface growth have been
identified, the  best-known of these are characterised
in terms of stochastic equations for the height profile $h=h(t,\vec{r})$
\BEQ \label{gl:KPZ-EW}
\left\{ \begin{array}{ll}
\partial_t h = \nu \nabla^2 h + \eta & \mbox{\rm ~~;~ Edwards-Wilkinson~~ {\sc ew}~ \cite{Edwa82}} \\
\partial_t h = \nu \nabla^2 h + \frac{\mu}{2}\left(\nabla h\right)^2
+ \eta & \mbox{\rm ~~;~ Kardar-Parisi-Zhang {\sc kpz} \cite{Kard86}}
\end{array} \right.
\EEQ
where $\nabla$ is the spatial gradient, $\eta$ is a centred gaussian white noise, with covariance
\BEQ
\left\langle \eta(t,\vec{r})\eta(t',\vec{r}')\right\rangle = 2\nu T \delta(t-t')\delta(\vec{r}-\vec{r}')
\EEQ
and $\nu,\mu,T$ are material-dependent constants.

While the exact solution of the {\sc ew}-equation is straightforward,
extracting the long-distance and/or long-time properties of interfaces
in the {\sc kpz}-class is considerably more difficult and several aspects of the problem still remain unresolved.
Remarkable progress has been achieved in  recent years on the exact solution of the {\sc kpz}-equation in $d=1$ dimension.
In particular, several spatial correlators have been found exactly and a
deep relationship of the probability distribution ${\cal P}(h)$ of the
fluctuation $h-\overline{h}$ with the extremal value statistics of the largest eigenvalue of random matrices has been derived,
see \cite{Sasa10,Cala11,Cala14,Imam12,Halp12,Halp13,Kell17}. Very remarkably, these mathematical results could be
confirmed experimentally, in several physically distinct systems \cite{Take11,Huer12,Take12,Take14,Yunk13,Halp14,Halp14b,Huer14,Atis14,Take17}.
Still, this impressive progress seems to rely on specific properties of the one-dimensional case.
Therefore, one might wonder if further classes of exactly solvable models of interface growth could be defined,
distinct from both the {\sc ew}- as well as the {\sc kpz}-universality class, and what physical insight the study of such models might provide.

%%++++++++++++++++++++++++++++++++++++++++++++++++++++++++++++++++++++++++++++++++
\begin{figure}[tb]
\centerline{\psfig{figure=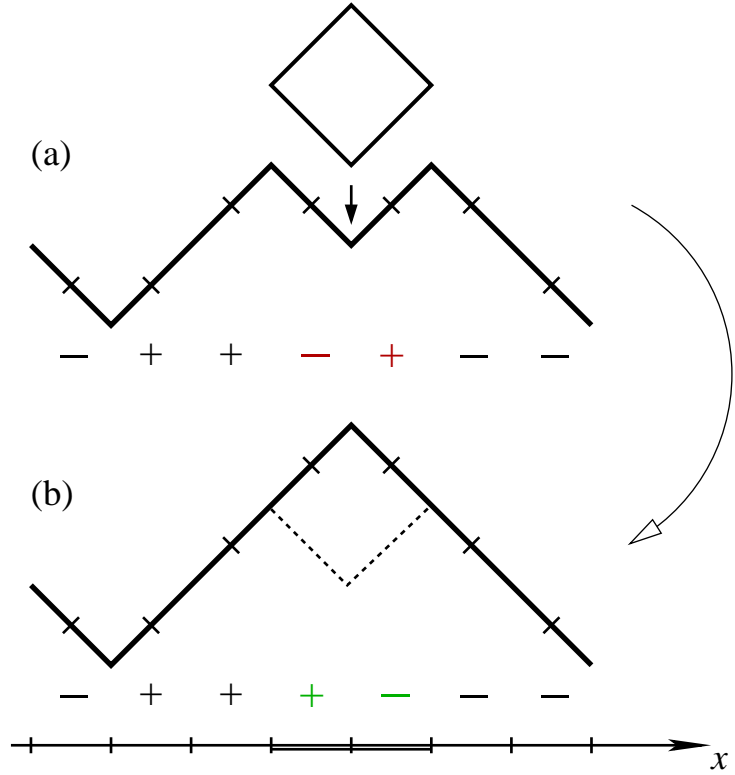,width=3.8in,clip=}}
\caption[fig1]{Illustration of the growth of an interface obeying the RSOS condition. In (a), the interface is shown
before the adsorption of a particle and in (b), the same interface is shown after the adsorption process.
Below the interfaces (a) and (b), the slopes $u_{j}=h_{j+\demi}-h_{j-\demi}$,
defined on the dual lattice $j\in\mathbb{Z}+\demi$ are indicated. The adsorption
process is described by a move $(-+)\to(+-)$ in terms of the slopes,
where the participating slopes are indicated in red before the
adsorption (a) and in green afterwards (b).  \label{fig1}
}
\end{figure}
%%++++++++++++++++++++++++++++++++++++++++++++++++++++++++++++++++++++++++++++++++

Indeed, a new class of models can be defined, with the help of some inspiration from the definition
of the well-studied {\em spherical model} of a ferromagnet \cite{Berl52,Lewi52}.
Therein, the traditional Ising spins $\sigma_i=\pm 1$, attached to the sites $i$
of a lattice with $\cal N$ sites, are replaced by
continuous spins $S_i\in\mathbb{R}$ and subject to the {\it `spherical constraint'}
$\sum_i S_i^2 = {\cal N}$. A conventional nearest-neighbour interaction leads
to an exactly solvable model, which undergoes a non-trivial phase transition in
$2<d<4$ dimensions \cite{Berl52,Joyc72}. The relaxational properties
can be likewise analysed exactly, see e.g. \cite{Ronc78,Coni94,Cugl94,Cugl95,Godr00b,Fusc02,Pico02,Dutt08,Fort12,Godr13}.
In order to identify an analogy with growing interfaces, we restrict here to
$d=1$ dimensions for simplicity. Consider
a lattice representation of the {\sc kpz}-class where the height differences
between two nearest neighbours obey the so-called {\sc rsos} constraint
$h_{i+1}(t)-h_i(t)=\pm 1$. It is well-established that in the continuum limit
this model is described by the {\sc kpz}-equation \cite{Bara95,Halp95,Taeu14},
see \cite{Bert97} for a rigorous derivation. The dynamic deposition rule is
sketched in figure~\ref{fig1}, which makes it clear that in this kind
of lattice model, the slopes $u_{i+1/2}(t):=h_{i+1}(t)-h_{i}(t)$
should be considered as the analogues of the Ising spins $\sigma_{i}=\pm 1$ in ferromagnets.
For the slopes, in the continuum limit, from
the {\sc kpz}-equation follows the (noisy) Burgers equation \cite{Burg74} (for a discrete analogue, see \cite{benN12})
\BEQ
\partial_t u = \nu \nabla^2 u + \mu u \nabla u + \nabla \eta
\EEQ

A `spherical model variant' of the {\sc kpz}-universality class now stipulates to
relax the {\sc rsos}-constraints $u_i=\pm 1$ to a `spherical constraint'
$\sum_i \langle u_i^2 \rangle = {\cal N}$ \cite{Henk15}.\footnote{An old observation by
Oono and Puri \cite{Oono88} gives additional motivation: treating the Allen-Cahn equation of phase-ordering,
after a quench to $T<T_c$, along the lines of the celebrated Ohta-Jasnow-Kawasaki approximation, but for a {\em finite} thickness of
the domain boundaries, leads to a kinetic equation in the universality class of the spherical model.}
However, for growing interfaces several equivalent descriptions can
give rise to several new models, which may or may not
be in the same universality class. Heuristically, the following possibilities may occur:

{\bf 1.} One may start from the Burgers equation and replace its non-linearity as follows
\BEQ \label{Arc1}
\partial_t u = \nu \nabla^2 u + \mu u \nabla u + \nabla \eta ~~\mapsto~~
\partial_t u = \nu \nabla^2 u + \mathfrak{z}(t) u + \nabla \eta
\EEQ
with a Lagrange multiplier $\mathfrak{z}(t) \sim \left\langle \nabla u\right\rangle$ which
might be seen as some kind of `averaged curvature' of the interface.
Its value is determined by the mean spherical constraint\footnote{In this section, the average $\langle\cdot\rangle$ is
understood to be taken over both the `thermal' as well as the `initial' noise.}
$\sum \langle u^2\rangle  = {\cal N}$. This is the {\em `first Arcetri model'},
defined\footnote{The name comes from the
location of the Galileo Galilei Institute of Physics, where this model was conceived.}
and analysed in \cite{Henk15}.\footnote{It can be shown that $\mathfrak{z}(t)\sim t^{-1}$
for sufficiently long times, whenever $T\leq T_c(d)$.}
In any dimension $d>0$, there is a `critical temperature' $T_c(d)>0$
such that long-range correlations build up for $T\leq T_c(d)$.
At the critical point $T=T_c(d)$, the interface is rough for $d<2$ and is smooth for $d>2$.
For $T<T_c(d)$, the interface is always rough. The model is also related to the gaps in the spectra of random matrices \cite{Fyod15} and to the
spherical spin glass \cite{Cugl95}.

{\bf 2.} An alternative way to treat the Burgers equations might proceed as follows
\BEQ \label{Arc2}
\partial_t u = \nu \nabla^2 u + \mu u \nabla u+ \nabla \eta ~~\mapsto~~
\partial_t u = \nu \nabla^2 u + \mathfrak{z}(t) \nabla u + \nabla \eta
\EEQ
where the Lagrange multiplier $\mathfrak{z}(t) \sim \left\langle u\right\rangle$
might now be viewed as some kind of `averaged slope'.
Its value is again determined by the constraint $\sum \langle u^2\rangle  = {\cal N}$.
This would define a {\em `second Arcetri model'}.

{\bf 3.} Finally, we might have started directly from the {\sc kpz} equation
\BEQ \label{Arc3}
\partial_t h = \nu \nabla^2 h + \demi \mu \left( \nabla h\right)^2 + \eta ~~\mapsto~~
\partial_t h = \nu \nabla^2 h + \mathfrak{z}(t) \nabla h +\eta
\EEQ
where $\mathfrak{z}(t) \sim \left\langle \nabla h\right\rangle$ might again be interpreted
as an `averaged slope' and will be found from a
constraint $\sum \langle\left( \nabla h\right)^2\rangle = {\cal N}$.
This would be a {\em `third Arcetri model'.}

However, such a simplistic procedure would lead to undesirable
properties of the height and slope profiles in the stationary state, as well as to internal inconsistencies.
We shall therefore reconsider this correspondence carefully in section~2,
where the precise definitions of the second  and third Arcetri model will be given.

In one spatial dimension, the slope profile has an interesting relationship with the dynamics of interacting particles.
To see this, write the slope as $u(t,r) = 1 - 2\vro(t,r)$,  where $\vro(t,r)$  denotes the particle-density at time $t\in\mathbb{R}_+$
and position $r\in\mathbb{R}$. In the {\sc kpz} universality class, when on the lattice the {\sc rsos}-constraint $u(t,r)=\pm 1$ holds,
denote by $\bullet$ an occupied site with $\vro=1$ and by $\circ$ an empty site with $\vro=0$. Then the only admissible reaction between
neighbouring sites is the directed jump $\bullet\circ\longrightarrow\circ\bullet$.
The stochastic process described by these interacting particles is
a {\em totally asymmetric exclusion process} ({\sc tasep}), see e.g. \cite{Ligg85,Gwa92,Daqu11,Mall15},
which is integrable via the Bethe ansatz.
Here, we are interested in the situation when the exact {\sc rsos}-constraint is relaxed to the mean `spherical constraint'
$\left\langle \sum_r u(t,r)^2 \right\rangle = {\cal N}$. In terms of the noise-averaged particle-density, this becomes
\BEQ \label{gl:rho}
\sum_r \left\langle \vro(t,r) \right\rangle = \sum_r \left\langle \vro(t,r)^2 \right\rangle
\EEQ
where the sums run over all sites of the lattice. Hence, on any site, neither $\langle\vro(t,r)\rangle$ nor
the difference $\langle \vro(t,r)\rangle -1$ can become
very large, since the spherical constraint prohibits the condensation of almost all particles onto
a very small number of sites.
In particular, if one takes a spatially translation-invariant initial condition,
then spatial translation-invariance is kept for all times.
Because of the constraint (\ref{gl:rho}), the average (position-independent) particle-density
\BEQ
\rho(t) := \frac{1}{\cal N}\sum_r \left\langle \vro(t,r) \right\rangle \geq 0
\EEQ
is always non-negative.
We point out that while the non-averaged density variable
$\vro(t,r)\in\mathbb{R}$ has no immediate physical interpretation,
the constraint (\ref{gl:rho}) guarantees that the measurable
{\em disorder-averaged observables} takes physically reasonable values.

%%--------------------------------------------------------------------------------------------------------
\begin{table}[tb]
\caption[tab1]{Non-equilibrium exponents, as defined in the text, for several universality classes of growing-interface models. The
{\sc Arcetri 1h} class at $T=T_c$ for $d>2$ is identical to the {\sc ew} class \cite{Edwa82,Roet06,Bust07,Igua09,Kell17}. 
For the {\sc Arcetri 3} class, there are three
distinct logarithmic sub-ageing scaling regimes, which are characterised by the value of the
logarithmic sub-ageing parameter $\vartheta$ as indicated. For $\vartheta<1$, the autoresponse function does not display scaling behaviour,
as indicated by {\sc dns} ({\bf d}oes {\bf n}ot {\bf s}cale). For empty entries, no estimate is known. 
The initial state is flat on average, with uncorrelated heights.
\label{tab1}}
{\small %\begin{center}
\begin{tabular}{|lc|ccccccl|} \hline
model & $d$ & $a$           & $b$           & $\lambda_C$ & $\lambda_R$ & $z$        & $\beta$       & Ref. \\ \hline
KPZ   & 1   & $-1/3$~~~     & $-2/3$~~~~~~~ & $1$         & $1$         & $3/2$~~    & $1/3$~~~      & {\footnotesize\cite{Kard86,Krec97,Kall99,Henk12}} \\
KPZ   & 2   &               & $-0.483$~~~~~ & $1.91(6)$   &             & $1.61(2)$~ & $0.241(1)$~~  & \cite{Halp14} \\
KPZ   & 2   & $0.30(1)$     & $-0.483(3)$~  & $1.97(3)$   & $2.04(3)$   & $1.61(2)$~ & $0.2415(15)$  & \cite{Odor14} \\
KPZ   & 2   & $0.24(1)$     & $-0.483(3)$~  & $1.97(3)$   & $2.00(6)$   & $1.61(2)$~ & $0.2415(15)$  & \cite{Kell16} \\
KPZ   & 2   & $0.24(1)$     & $-0.4828(4)$  & $1.98(5)$   & $2.00(6)$   & $1.611(3)$ & $0.2414(2)$~~ & \cite{Kell17} \\
      &     &               &               & $2.01(2)$   &             &            &               &               \\[0.25truecm]
{\sc Arcetri 1h} $T=T_c$ & $<2$
            & $\freins-1$   & $\freins-1$   & $\frdrei-1$ & $\frdrei-1$ & $2$        & $\frac{1}{4}(2-d)$ & \cite{Henk15} \\[0.12truecm]
{\sc Arcetri 1h} $T=T_c$ & $>2$
            & $\freins-1$   & $\freins-1$   & $d$         & $d$         & $2$        & $0$          & \cite{Henk15} \\[0.12truecm]
{\sc Arcetri 1h} $T<T_c$ & $d$
            & $\freins-1$   & $-1$          & $\freins-1$ & $\freins-1$ & $2$        & $\demi$      & \cite{Henk15} \\[0.25truecm]
{\sc Arcetri 3} \hfill $T=0$   & $1$
            & $-\demi$   & $0$              & $0$         & $1$         & $2$        & $0$          & $\vartheta>1$ \\[0.10truecm]
{\sc Arcetri 3} \hfill $T=0$   & $1$
            & {\sc dns}        & $0$        & $0$         & {\sc dns}   & $2$        & $0$          & $\demi<\vartheta<1$ \\[0.10truecm]
{\sc Arcetri 3} \hfill $T=0$   & $1$
            & {\sc dns}        & $0$        & $\infty$    & {\sc dns}   & $2$        & $0$          & $\vartheta=\demi$ \\ \hline
\end{tabular}%\end{center}
}
\end{table}
%%--------------------------------------------------------------------------------------------------------

The long-time non-equilibrium relaxation behaviour is analysed as follows.
In models of interface growth, one usually starts from a flat, horizontal interface with uncorrelated heights
\cite{Bara95,Halp95,Roet06,Henk12,Halp14,Odor14,Henk15,Kell16,Kell17,Kell17b}. One then studies the average height
$\langle h(t,r)\rangle\sim t^{\beta}$, the interface width
$w^2(t)= \langle \left(h(t,r) - \langle h(t,r)\rangle \right)^2 \rangle \sim t^{2\beta}$, and the
two-time height autocorrelator and auto-response of the height with respect to a change in the height
\BEA
C(t,s) &:=& \langle \left( h(t,r) - \langle h(t,r)\rangle \right)\left( h(s,r) - \langle h(s,r)\rangle \right) \rangle
\:=\: s^{-b} f_C\left(\frac{t}{s}\right) \label{scalingCh}
\\
R(t,s) &:=& \left. \frac{\delta \langle h(t,r) \rangle}{\delta j(s,r)}\right|_{j=0} \:=\: s^{-1-a} f_R\left(\frac{t}{s}\right) \label{scalingRh}
\EEA
The scaling forms \cite{Fami85} used here are those of {\em simple ageing} 
and apply in the long-time limit $t,s\to\infty$ with $y=t/s$ being kept fixed.
The scaling functions are expected to have the asymptotic behaviour%, for $y\gg 1$ 
\BEQ \label{echelle1.11}
f_{C}(y) \stackrel{y\gg 1}{\sim} y^{-\lambda_{C}/z} \;\; , \;\; 
f_{R}(y) \stackrel{y\gg 1}{\sim} y^{-\lambda_{R}/z}
\EEQ
where $z$ is the dynamical exponent. From
these relations the exponents $\beta$ and $a,b,\lambda_C,\lambda_R$ are defined. In table~\ref{tab1}, 
some values of these exponents are collected.\footnote{The $2D$ {\sc kpz} universality class is realised by the octahedron model \cite{Odor14}. 
For height correlators and responses, the results of the random sequential (RS) update and of the two-sublattice stochastic dynamics (SCA) update 
are consistent, confirming the expected universality ($\lambda_{C,\mbox{\rm\tiny RS}}=1.98(5)$, $\lambda_{C,\mbox{\rm\tiny SCA}}=2.01(2)$) \cite{Kell17}.
Comparison with the recent result $z=1.613(2)$ \cite{Pagn15} gives an {\it a posteriori} indication of the presently achieved numerical precision.} 
Starting from the Langevin equation (\ref{Arc1}) of the first Arcetri model, formulated in terms of the the slopes $u$ and using $u=\nabla h$,
an analogous Langevin equation for the heights $h$ is found, if only the spherical constraint is now written as $\sum \langle(\nabla h)^2\rangle={\cal N}$.
In what follows, we shall call this the {\it {\sc Arcetri 1h} model}. Its relaxational behaviour undergoes (simple) ageing for both $T=T_c$ and for
$T<T_c$, in agreement with the expected scaling forms (\ref{scalingCh},\ref{scalingRh}).
In appendix~A, we briefly outline how to find the exponents. 
Logarithmic sub-scaling exponents \cite{Kenn06} in $w(t)$ of the third Arcetri model are discussed in section~4. 

The main focus of this work will be on defining (see section~2 for the precise definitions) 
and analysing the `second' and the `third' Arcetri models. At temperature $T=0$, 
we shall see that the simple ageing behaviour of eqs.~(\ref{scalingCh},\ref{scalingRh},\ref{echelle1.11}) does not apply. 
Rather, we shall find a `{\it logarithmic sub-ageing}' behaviour,\footnote{{\it Sub-ageing behaviour} is defined by the scaling variable 
$y-1:=\frac{\kappa_2(t-s)}{(\kappa_2 s)^{\mu}}$, where $0<\mu<1$ is the {\it sub-ageing exponent} and $\mu=1$ gives back simple ageing \cite{Cugl03,Vinc07}.
See \cite[Tab 1.2]{Henk10} for a list of experimentally measured values of $\mu$. 
A basic rigorous inequality excludes the case $\mu>1$ (`super-ageing') \cite{Kurc02}.}
in the scaling limit where both times $t,s\to\infty$, but such that the scaling variable $y$ of two-time scaling 
\BEQ \label{eq:sous-vieil}
y-1 := \frac{t-s}{s} \ln^{\vartheta} \kappa_2 s 
\EEQ
is being kept fixed ($\kappa_2$ is a model-dependent constant). It turns out that several types of logarithmic sub-ageing exist for the Arcetri models, 
which are characterised by different values of the logarithmic sub-ageing exponent $\vartheta>0$.\footnote{For $\vartheta=0$, one is back to simple ageing} 
With the scaling variable (\ref{eq:sous-vieil}), the asymptotic scaling forms (\ref{echelle1.11}) often remain applicable and the corresponding exponent
values are quoted in tables~\ref{tab1} and~\ref{tab2}. Logarithmic sub-ageing arises from the presence of several time-dependent length scales, which differ
by factors logarithmic in time, a phenomenon also referred to as {\it multiscaling} \cite{Coni94}. 
If the autocorrelator scaling function $f_C(y)$ decays with $y$ faster than a power-law 
(exponentially or stretched exponentially), the value $\lambda_C=\infty$ is quoted. See section~5 for a fuller discussion. 

%%--------------------------------------------------------------------------------------------------------
\begin{table}[tb]
\caption[tab2]{Non-equilibrium exponents for several universality classes of lattice-gas models, 
where `octa' stands for `octahedron model', RS for random sequential update and SCA for two-sub-lattice stochastic dynamics. 
The model realisations in the {\sc kpz} and {\sc ew} universality classes are indicated. 
For the {\sc Arcetri 2} and conserved spherical classes, there
are three logarithmic sub-ageing scaling regimes which are characterised by the value of the
logarithmic sub-ageing parameter $\vartheta$ is indicated. 
For $\vartheta<1$, the autoresponse functions do not display scaling behaviour, indicated by {\sc dns}.
For empty entries, there is no estimate. 
The initial state has an average particle density $\vro=\demi$ and uncorrelated particles.
\label{tab2}}
{\small %\begin{center}
\begin{tabular}{|lc|cccccccl|} \hline
model                        & $d$ & $a$  &$a_{\cal R}$  & $b$      & $\lambda_C$ & $\lambda_R$ & $\lambda_{\cal R}$ & $z$        & Ref. \\ \hline
TASEP                        & 1   &      &              & $2/3$    & $3$         &             &                    & $3/2$      & \cite{Daqu11} \\[0.10truecm]
octa RS \hfill  {\fn\sc kpz} & 2   &      &              &$0.76(2)$ & $3.8(2)$    &             &                    & $1.611(3)$ & \cite{Kell17} \\[0.10truecm]
octa SCA \hfill {\fn\sc kpz} & 2   &      &              &          & $1.25(2)$   &             &                    & $1.611(3)$ & \cite{Kell17} \\[0.25truecm]
octa SCA \hfill {\fn\sc ew}  & 2   &      &              &$1.1(2)$  & $\approx 4$ &             &                    & $2$       & \cite{Kell17} \\[0.10truecm]
octa RS \hfill  {\fn\sc ew}  & 2   &      &              &$1.1(2)$  & $1.4(4)$    &             &                    & $2$       & \cite{Kell17} \\[0.25truecm]
{\sc Arcetri 1u} $T=T_c$     & $d$ &$d/2$ & $d/2+1$      & $d/2$    & $d+2$       & $d$         & $d+2$              & $2$       &  \\
{\sc Arcetri 1u} $T<T_c$     & $d$ &$d/2-1$ & $d/2$      & $0$      & $d/2$       & $d/2$       & $d/2+2$            & $2$       &  \\[0.25truecm]
{\sc Arcetri 2} \hfill $T=0$ & $1$ &$-1/2$  & $0$        & $0$      & $0$         & $1$         & $1$                & $2$       & $\vartheta>1$ \\[0.10truecm]
{\sc Arcetri 2} \hfill $T=0$ & $1$ &{\sc dns} &{\sc dns} & $0$      & $0$         & {\sc dns}   &                    & $2$  & $\demi<\vartheta<1$ \\[0.10truecm]
{\sc Arcetri 2} \hfill $T=0$ & $1$ &{\sc dns} &{\sc dns} & $0$ & $\infty$    & {\sc dns}   &                    & $2$       & $\vartheta=\demi$ \\[0.25truecm]
spherical \hfill $T=0$       & $d$ &$(d-2)/4$ &          & $0$ & $0$         & $d+2$       &                    & $4$       & $\vartheta>1$ \\[0.10truecm]
spherical \hfill $T=0$       & $d$ &{\sc dns} &{\sc dns} & $0$ & $0$         & {\sc dns}   & {\sc dns}          & $4$       & $\demi<\vartheta<1$ \\[0.10truecm]
spherical \hfill $T=0$       & $d$ &{\sc dns} &{\sc dns} & $0$ & $\infty$    & {\sc dns}   & {\sc dns}   & $4$ & $\vartheta=\demi$ \hfill ~~~~\cite{Bert00} 
\\ \hline
\end{tabular} %\end{center}
}
\end{table}
%%--------------------------------------------------------------------------------------------------------

Analogously, if one considers a system of interacting particles, one usually assumes an initial state of uncorrelated particles
(uncorrelated, flat slopes $\langle u(t,r)\rangle=0$, in the present terminology) with an average particle density
$\vro=\langle \vro(t,r)\rangle=\demi$, equivalent to a vanishing initial slope \cite{Daqu11,Kell17}. 
One considers the two-time slope (connected) auto-correlator $C(t,s)$, which is related to the density-density autocorrelator,
and the linear auto-responses $R(t,s)$, ${\cal R}(t,s)$ of the slope with respect to a change $k=\nabla j$ in the slope or a change $j$
in the height, respectively
\BEA
C(t,s) &:=& \langle  u(t,r) u(s,r) \rangle \:=\: 4  \left\langle \left(\vro(t,r) -\demi\right)\left(\vro(s,r)-\demi\right) \right\rangle
\:=\: s^{-b} f_C\left(\frac{t}{s}\right) \label{1.11}
\\
R(t,s) &:=& \left. \frac{\delta \langle u(t,r) \rangle}{\delta k(s,r)}\right|_{k=0} \:=\: s^{-1-a} f_R\left(\frac{t}{s}\right)
\\
{\cal R}(t,s) &:=& \left. \frac{\delta \langle u(t,r) \rangle}{\delta j(s,r)}\right|_{j=0} \:=\: s^{-1-a_{\cal R}} f_{\cal R}\left(\frac{t}{s}\right)
\EEA
along with the expected behaviour of simple ageing in the scaling limit. Eq.~(\ref{echelle1.11}) applies again and analogously, one anticipates
$f_{{\cal R}}(y)\sim y^{-\lambda_{{\cal R}}/z}$, for $y\gg 1$.  
Considering numerical simulations of the $2D$ octahedron model, however, it appears that for the slope correlations 
the two update schemes RS and SCA lead to different values of the 
autocorrelation exponent -- and this for model realisation both in the {\sc kpz} as well as in the {\sc ew} universality classes \cite{Kell17}. 
The first Arcetri model with initially uncorrelated slopes will be called the
{\it {\sc Arcetri 1U} model}. It is suggestive to compare the corresponding exponent values with those of the {\sc ew} universality class. 
Some values of these exponents are listed in table~\ref{tab2},
see appendix~A for the outline of the calculations in the {\sc Arcetri 1u} model.
We also include results from the spherical model with a conserved order parameter (`model B'), at $T=0$ \cite{Coni94,Bert00}. 
It also becomes apparent how much less is known about responses of the slope variables than for the height variables.

This work is organised as follows: in section~2, the second and third Arcetri model are carefully defined.
Since the first Arcetri model was already studied \cite{Henk15},
we merely outline its treatment in appendix~A and quote the results in tables~\ref{tab1} and~\ref{tab2}, where the two possible
interpretations are taken into account.
Section~3 explains the solution of the second and third models.
The explicit spherical constraints and a closed form for correlators and responses are derived.
In section~4, the asymptotic analysis at temperature $T=0$ and the emergence of the different types of
logarithmic sub-ageing in the second and third models is presented.
We conclude in section~5 with a detailed presentation of the kinetic phase diagram and the various scales on
which different aspects of logarithmic sub-ageing occur.
Technical calculations are treated in several appendices.
Appendix~A contains a short summary of the first model, both for an interface and for a lattice gas.
Appendices~B and~C derive the various distinct sub-ageing scaling forms
of correlators and responses, respectively.
Several mathematical identities are derived in appendices~D and~E
and some basics of discrete cosine- and sine transformations are collected in appendix~F.

%%%%%%%%%%%%%%%%%%%%%%%%%%%%%%%%%%%%%%%%%%%%%%%%%%%%%%%%%%%%%%%%%%%%%%%%%%%%%%%%
\section{The second and third Arcetri models}
%%%%%%%%%%%%%%%%%%%%%%%%%%%%%%%%%%%%%%%%%%%%%%%%%%%%%%%%%%%%%%%%%%%%%%%%%%%%%%%%

%%%%%%%%=======================================================================
\subsection{Preliminaries}
%%%%%%%%=======================================================================

Why are the equations (\ref{Arc2},\ref{Arc3}) physically unsatisfactory~? In order to understand this,
and in consequence the necessity for a better definition of the models,
consider for a moment the behaviour of the stationary profiles,
as they would follow from eqs.~(\ref{Arc1},\ref{Arc2},\ref{Arc3}).
Let $\mathfrak{z}_{\infty}$ denote the stationary value of the Lagrange multiplier.
Then the noise-averaged slope profile of the first Arcetri model (\ref{Arc1}) is oscillatory
$u_{\rm stat}(r)\sim \cos \left(r/\lambda + \vph_0\right)$,
with the finite wave-length $\lambda= \sqrt{\nu/\mathfrak{z}_{\infty}\,}$, as one would have expected.
On the other hand, eq.~(\ref{Arc2}) would produce a spatially strongly variable stationary slope profile
$u_{\rm stat}(r) \sim \exp\left( - r/r_0\right)$, with a finite length-scale $r_0 = \nu/\mathfrak{z}_{\infty}$.
Finally, eq.~(\ref{Arc3}) gives an analogous result for the stationary height profile.
This is in apparent contradiction with the expectation of essentially flat profiles,
both for the height as well as the slope.

%%%%%%%%=======================================================================
\subsection{Definition of the second Arcetri model}
%%%%%%%%=======================================================================

How can one formulate a physically sensible `spherical model variant' of the Burgers equation~?
Begin with a decomposition of the slope profile $u(t,r)$, with $r\in\mathbb{R}$, into its even and odd parts
\BEQ
u(t,r) = a(t,r) + b(t,r)
\EEQ
where
\BEA
a(t,r) &:=& \demi \left( u(t,r) + u(t,-r) \right) \:=\: ~~a(t,-r) \mbox{\rm ~~ even} \nonumber \\
b(t,r) &:=& \demi \left( u(t,r) - u(t,-r) \right) \:=\: -b(t,-r) \mbox{\rm ~~~ odd}
\EEA
For definiteness, we shall formulate the defining equations of motion of the {\em second Arcetri model}
on a periodic chain of $N$ sites. They read
\BEA
\partial_t a_n(t) &=& \nu \left( a_{n+1}(t) + a_{n-1}(t) -2 a_n(t) \right) + \demi \mathfrak{z}(t) \left( b_{n+1}(t) - b_{n-1}(t) \right)
+ \demi \left( \eta_{n+1}^{-}(t) - \eta_{n-1}^{-}(t) \right)
\nonumber \\
& & \label{Arcetri2} \\
\partial_t b_n(t) &=& \nu \left( b_{n+1}(t) + b_{n-1}(t) -2 b_n(t) \right) - \demi \mathfrak{z}(t) \left( a_{n+1}(t) - a_{n-1}(t) \right)
- \demi \left( \eta_{n+1}^{+}(t) - \eta_{n-1}^{+}(t) \right)
\nonumber
\EEA
%\textcolor{blue}{\tt des d\'efs. des r\'eponses, il faut fixer $\textcolor{red}{\eps=-1}$.}
where $\eta_n^{\pm}(t) := \demi \left( \eta_n(t) \pm \eta_{N-n}(t) \right)$
is the parity-symmetrised and -antisymmetrised white noise $\eta_n(t)$, with  the moments
\BEQ
\left\langle \eta_n(t) \right\rangle = 0 \;\; , \;\;
\left\langle \eta_n(t) \eta_m(t') \right\rangle = 2\nu T \delta_{n,m} \delta(t-t')
\EEQ
Hence one  has the (anti-)symmetrised noise correlators
\BEQ \label{bruit_symm}
\left\langle \eta_n^{\pm}(t) \right\rangle = 0 \;\; , \;\;
\left\langle \eta_n^{\pm}(t) \eta_m^{\pm}(t') \right\rangle =
\nu T \delta(t-t')\left[ \delta_{n,m} \pm \delta_{n,N-m} \right] \;\; , \;\;
\left\langle \eta_n^{\pm}(t) \eta_m^{\mp}(t') \right\rangle =0
\EEQ
(clearly, the indices $n,m$ are to be taken {\it modulo} $N$).
The second Arcetri model will be considered as a variant of the Burgers equation and its associated {\sc tasep}.
Therefore, a natural choice of initial conditions is to admit initially uncorrelated slopes,
distributed according to a gaussian, and with the moments
\BEA
\left\langle\!\left\langle a_n(0) \right\rangle\!\right\rangle &=& \left\langle\!\left\langle b_n(0) \right\rangle\!\right\rangle = 0
\hspace{1.0truecm}\;\; , \;\;
\left\langle\!\left\langle a_n(0)b_m(0) \right\rangle\!\right\rangle = 0
\nonumber \\
\left\langle\!\left\langle a_n(0)a_m(0) \right\rangle\!\right\rangle &=& \demi\left( \delta_{n,m} + \delta_{n,N-m} \right)
\;\; , \;\;
\left\langle\!\left\langle b_n(0)b_m(0) \right\rangle\!\right\rangle = \demi\left( \delta_{n,m} - \delta_{n,N-m} \right)
\label{Arc2_initial}
%une amplitude est normalis\'ee: $U_1^2=1$
\EEA
The Lagrange multiplier $\mathfrak{z}(t)$ is determined from the mean spherical constraint on the slopes
\BEQ \label{Arc2_contrainte}
\left\langle\!\!\!\left\langle \left\langle \sum_{n=1}^{N}
\left( a_n(t) + b_n(t) \right)^2 \right\rangle \right\rangle\!\!\!\right\rangle  = N
\EEQ
which is averaged over both sources of noise present in the model, as indicated by the brackets
$\langle .\rangle$ for the average over $\eta$ and
$\langle\!\langle .\rangle\!\rangle$ for the average over the initial conditions.
We stress that the even and odd parts are treated in a slightly different way.
In this way, two essential properties of the Burgers equation,
namely (i) the conservation law and (ii) the non-invariance under the parity
transformation $x\mapsto -x$ \cite{Burg74,benN12} are kept. The initial conditions (\ref{Arc2_initial}) are natural if one wishes
to interpret the slope $u(t,r)=1-2\vro(t,r)$ in terms of the density $\vro(t,r)$ of a model of interacting particles, with
the average density $\vro = \left\langle\!\left\langle\,\langle \vro(t,r)\rangle\, \right\rangle\!\right\rangle=\demi$.

Eqs.~(\ref{Arcetri2},\ref{Arc2_initial},\ref{Arc2_contrainte}), together with the noise correlator (\ref{bruit_symm}), define the
{\bf second Arcetri model}.

Formally, one might also arrive at these equations by introducing a complex velocity $u=a+\II b$ into the modification (\ref{Arc2})
of the Burgers equation, with a complex Lagrange multiplier $\mathfrak{z}(t)= \mathfrak{z}_1(t)-\II \mathfrak{z}_2(t)$ and a complex
noise $\eta(t,r) = \II \left( \eta^{+}(t,r) - \II \eta^{-}(t,r)\right)$. Separating into real and imaginary parts, this would give
\BEA
\partial_t a &=& \nu \partial_r^2 a +\mathfrak{z}_1 \partial_r a + \mathfrak{z}_2 \partial_r b +\partial_r \eta^{-}
\nonumber \\
\partial_t b &=& \nu \partial_r^2 b +\mathfrak{z}_1 \partial_r b - \mathfrak{z}_2 \partial_r a -\partial_r \eta^{+}
\nonumber
\EEA
%\textcolor{blue}{\tt on a fix\'e $\textcolor{red}{\eps=- 1}$.}
Only if one chooses $\mathfrak{z}_1=0$, one obtains an oscillatory equation
$\nu^2 p'' = -\mathfrak{z}_{2,\infty}^2 p$ for the derivative $p :=\lim_{t\to\infty}\partial_r \langle a\rangle$
of the noise-averaged stationary slope, and similarly for $q := \lim_{t\to\infty}\partial_r \langle b\rangle$.
The effect of this formally `imaginary' Lagrange multiplier is included in the equations of motion (\ref{Arcetri2}).
%\textcolor{blue}{\tt cette proc\'edure donne directement \textcolor{red}{$\eps=-1$}, dans notre vieille notation.
%Ceci est d\'ej\`a admis ici pour le 2e Arcetri.}

Conservation laws become explicit by rewriting the complex equations of motion (\ref{Arc2})
\BEA
\partial_t (a+\II b) &=& \partial_r^2 (a+\II b) + \mathfrak{z}(t) \partial_r(a+\II b) +\partial_r(\eta^{-}-\II\eta^{+})
\nonumber \\
&=& \partial_r \left[ (\partial_r -\II\mathfrak{z}(t)) (a+\II b) -\II(\eta^{+}+\II\eta^{-}) \right]
\nonumber
\EEA
in the form of a continuity equation. Using $u=a+\II b$ and its formal complex conjugate $u^* = a -\II b$, along with
$\zeta :=\eta^{+}+\II\eta^{-}$ and $\zeta^* = \eta^{+}-\II\eta^{-}$, we have the pair of equations
\BEQ \label{2.8}
\partial_t u = \partial_r\left[ (\partial_r -\II\mathfrak{z}(t))u -\II\zeta \right] \;\; , \;\;
\partial_t u^* = \partial_r\left[ (\partial_r +\II\mathfrak{z}(t))u^* +\II\zeta^* \right]
\EEQ
and identify the densities $j=(\partial_r -\II\mathfrak{z}(t))u -\II\zeta$ and $j^*=(\partial_r +\II\mathfrak{z}(t))u^* +\II\zeta^*$ of the
conserved currents, such that the `conserved charges' $U=\int_{\mathbb{R}}\!\D r\: u(t,r)$ and $U^*=\int_{\mathbb{R}}\!\D r\: u^*(t,r)$ are time-independent,
viz. $\partial_t U = \partial_t U^*=0$.

%%%%%%%%=======================================================================
\subsection{Definition of the third Arcetri model}
%%%%%%%%=======================================================================

Analogously, for the third Arcetri model we start from the height profile $h(t,r)$, decomposed into even and odd parts
\BEQ
h(t,r) = a(t,r) + b(t,r)
\EEQ
and write down the defining equations of motion (on a discrete chain of $N$ sites)
\BEA
\partial_t a_n(t) &=& \nu \left( a_{n+1}(t) + a_{n-1}(t) -2 a_n(t) \right) + \demi \mathfrak{z}(t) \left( b_{n+1}(t) - b_{n-1}(t) \right)
+ \eta_n^+(t)
\nonumber \\
& & \label{Arcetri3} \\
\partial_t b_n(t) &=& \nu \left( b_{n+1}(t) + b_{n-1}(t) -2 b_n(t) \right) - \demi \mathfrak{z}(t) \left( a_{n+1}(t) - a_{n-1}(t) \right)
+ %\textcolor{red}{\eps}
\eta_n^{-}(t)
\nonumber
\EEA
with the symmetrised noise (\ref{bruit_symm}).
In this physical context, it appears natural to use initially uncorrelated gaussian slopes
\BEA
\left\langle\!\left\langle a_n(0) \right\rangle\!\right\rangle &=& H_0
\;\;\;\; , \;\;\;\;
\left\langle\!\left\langle b_n(0) \right\rangle\!\right\rangle \:=\: 0
\;\; , \;\;
\left\langle\!\left\langle a_n(0)b_m(0) \right\rangle\!\right\rangle \:=\: 0
\nonumber \\
 \left\langle\!\left\langle a_n(0)a_m(0) \right\rangle\!\right\rangle_c &=& \demi H_1 \left( \delta_{n,m} + \delta_{n,N-m} \right)
\;\; , \;\;
\left\langle\!\left\langle b_n(0)b_m(0) \right\rangle\!\right\rangle = \demi H_1 \left( \delta_{n,m} - \delta_{n,N-m} \right) ~~~~
\label{Arc3_initial}
\EEA
%\textcolor{blue}{\tt en analogie avec {\sc Arcetri 1h}, la forme d\'etaill\'ee a \'et\'e  v\'erifi\'ee.}
The Lagrange multiplier $\mathfrak{z}(t)$ is found from the mean spherical constraint on the slopes
\BEQ \label{Arc3_contrainte}
\left\langle\!\!\!\left\langle \left\langle \sum_{n=1}^{N}
\left( \nabla a_n(t) + \nabla b_n(t) \right)^2 \right\rangle \right\rangle\!\!\!\right\rangle  = N
\EEQ
where $\nabla f_n=\demi\left( f_{n+1} - f_{n-1}\right)$ is the symmetrised spatial difference. The initial conditions (\ref{Arc3_initial}) are
natural for an interpretation of $h(t,r)$ as the height of a growing and fluctuating interface, which is flat on average.

Eqs.~(\ref{Arcetri3},\ref{Arc3_initial},\ref{Arc3_contrainte}), together with the noise correlator (\ref{bruit_symm}), define the
{\bf third Arcetri model}.

Formally, one might obtain this from the modified {\sc kpz} equation (\ref{Arc3}) by introducing a complex height $h=a+\II b$,
a complex Lagrange multiplier $\mathfrak{z}(t) = \mathfrak{z}_1(t)-\II\mathfrak{z}_2(t)$
and a complex noise $\eta(t,r)=\eta^+(t,r) +\II\eta^{-}(t,r)$. As before, only if one chooses $\mathfrak{z}_1=0$, the derivative
$p := \lim_{t\to\infty}\partial_r \langle a\rangle$ of the stationary height obeys an oscillatory equation
$\nu^2 p'' = -\mathfrak{z}_{2,\infty}^2p$.
%\textcolor{blue}{\tt cette proc\'edure donne directement \textcolor{red}{$\eps=+1$}.}
Because of the `non-conserved' noise, there are no obvious conservation laws, for $T\ne 0$.

All definitions were only made explicit in $d=1$ spatial dimensions. Eventual extensions to $d>1$ are left for future work.

%%%%%%%%%%%%%%%%%%%%%%%%%%%%%%%%%%%%%%%%%%%%%%%%%%%%%%%%%%%%%%%%%%%%%%%%%%%%%%%%
\section{Solution}
%%%%%%%%%%%%%%%%%%%%%%%%%%%%%%%%%%%%%%%%%%%%%%%%%%%%%%%%%%%%%%%%%%%%%%%%%%%%%%%%

We begin our discussion with the second Arcetri model.
The treatment of the third Arcetri model being fairly analogous, we shall simply
quote the relevant results in section 3.4.

%%%%%%%%=======================================================================
\subsection{Second model: General form}
%%%%%%%%=======================================================================

The first step to the solution of eqs.~(\ref{Arcetri2}) proceeds via Fourier-transform,
but we must take into account the specific parity of the
$a_n$ and $b_n$. Therefore, we use the representation in terms of discrete cosine- and sine-transforms,
\BEQ
a_n(t) = \frac{1}{N}  \sum_{k=0}^{N-1} \cos\left(\frac{2\pi}{N} kn\right) \wht{a}(t,k)
\;\; , \;\;
b_n(t) = \frac{1}{N}  \sum_{k=0}^{N-1} \sin\left(\frac{2\pi}{N} kn\right) \wht{b}(t,k)
\EEQ
see appendix~F for details. Using eqs.~(\ref{Z3},\ref{Z4},\ref{Z5},\ref{Z6}), the equations of motion turn into
\BEA
\partial_t \wht{a}(t,k) &=& -2\nu \left( 1 - \cos{\small\left(\frac{2\pi}{N}k\right)}\right)\wht{a}(t,k)
+ \mathfrak{z}(t)\sin{\small\left(\frac{2\pi}{N}k\right)}\wht{b}(t,k)
+ \sin{\small\left(\frac{2\pi}{N}k\right)} \wht{\eta}^{-}(t,k)
\nonumber \\
\partial_t \wht{b}(t,k) &=& -2\nu \left( 1 - \cos{\small\left(\frac{2\pi}{N}k\right)}\right)\wht{b}(t,k)
+ \mathfrak{z}(t)\sin{\small\left(\frac{2\pi}{N}k\right)}\wht{a}(t,k)
+ \sin{\small\left(\frac{2\pi}{N}k\right)} \wht{\eta}^{+}(t,k)~~~~~~~
\label{3.2}
\EEA
Although we shall use the same notation for both cosine- and sine-transforms,
the parity must be taken into account for the inverse transformation.
We shall use the short-hands
\BEQ \label{eq:abbrevs}
\omega(k) := 1 - \cos\left(\frac{2\pi}{N}k\right) \;\; , \;\; \lambda(k) := \sin\left(\frac{2\pi}{N}k\right) \;\; , \;\;
Z(t) := \int_0^t \!\D\tau\: \mathfrak{z}(\tau)
\EEQ
Later,  when taking the continuum limit, it will be enough to simply replace
$\omega(k) \to 1-\cos k$ and $\lambda(k) \to \sin  k$, and to consider
$k\in (-\pi,\pi)$ instead of $k=0,1,\ldots,N-1$  on the chain.

The above equations (\ref{3.2}) are decoupled by going over to the combinations
$\wht{f}_{\pm}(t,k) := \wht{a}(t,k) \pm \wht{b}(t,k)$, which obey the equations
\BEQ
\partial_t \wht{f}_{\pm}(t,k) = \left( -2\nu \omega(k) \pm \lambda(k) \mathfrak{z}(t) \right) \wht{f}_{\pm}(t,k)
+\lambda(k) \left( \wht{\eta}^{-}(t,k) \pm\wht{\eta}^+(t,k)\right)
\EEQ
with the solutions
\BEA
\wht{f}_{\pm}(t,k) &=& \wht{f}_{\pm,00}(k)\: \exp\left[ -2\nu\omega(k)t \pm \lambda(k) Z(t)\right]
\\
& & + \int_0^t \!\D\tau\: \lambda(k) \left( \wht{\eta}^{-}(\tau,k)\pm\wht{\eta}^{+}(\tau,k)\right)
\exp\left[-2\nu\omega(k)(t-\tau) \pm \lambda(k) \left( Z(t) - Z(\tau)\right)\right] ~~~~ \nonumber
\EEA
and where the functions $\wht{f}_{\pm,00}$ are to be found from the initial conditions.
Going back to the parity eigenstates, using that $\wht{a} = \demi\left( \wht{f}_+ + \wht{f}_-\right)$ and
$\wht{b} = \demi\left( \wht{f}_+ - \wht{f}_-\right)$, we have explicitly
\BEA
\wht{a}(t,k) &=& \demi \left[ \wht{f}_{+,00}(k)\: e^{-2\nu\omega(k) t+\lambda(k) Z(t)}
+ \wht{f}_{-,00}(k)\: e^{-2\nu\omega(k) t-\lambda(k) Z(t)} \right]
\nonumber \\
& & +\demi \int_0^t \!\D\tau\, \lambda(k) \left[ \left( \wht{\eta}^{-}(\tau,k)+\wht{\eta}^{+}(\tau,k)\right)
e^{-2\nu\omega(k)(t-\tau) + \lambda(k) \left( Z(t) - Z(\tau)\right)} \right. \nonumber \\
& & \left. \hspace{2.2truecm}+ \left( \wht{\eta}^{-}(\tau,k)-\wht{\eta}^{+}(\tau,k)\right)
e^{-2\nu\omega(k)(t-\tau) - \lambda(k) \left( Z(t) - Z(\tau)\right)} \right]
\label{3.6} \\
\wht{b}(t,k) &=& \demi \left[ \wht{f}_{+,00}(k)\: e^{-2\nu\omega(k) t+\lambda(k) Z(t)}
- \wht{f}_{-,00}(k)\: e^{-2\nu\omega(k) t-\lambda(k) Z(t)} \right]
\nonumber \\
& & +\demi \int_0^t \!\D\tau\, \lambda(k) \left[ \left( \wht{\eta}^{-}(\tau,k)+\wht{\eta}^{+}(\tau,k)\right)
e^{-2\nu\omega(k)(t-\tau) + \lambda(k) \left( Z(t) - Z(\tau)\right)} \right. \nonumber \\
& & \left. \hspace{2.2truecm}- \left( \wht{\eta}^{-}(\tau,k)-\wht{\eta}^{+}(\tau,k)\right)
e^{-2\nu\omega(k)(t-\tau) - \lambda(k) \left( Z(t) - Z(\tau)\right)} \right]
\label{3.7}
\EEA
and a cosine or sine transformation, respectively, will bring back $a_n(t)$ and $b_n(t)$.
For the chosen initial conditions, we simply have
$\left\langle\!\!\!\left\langle \left\langle \wht{a}(t,k)\right\rangle \right\rangle\!\!\!\right\rangle
= \langle\!\!\!\langle \langle \wht{b}(t,k) \rangle \rangle\!\!\!\rangle =0$ which implies in turn
$\left\langle\!\!\!\left\langle \left\langle a_n(t)\right\rangle \right\rangle\!\!\!\right\rangle
= \left\langle\!\!\!\left\langle \left\langle b_n(t)\right\rangle \right\rangle\!\!\!\right\rangle =0$,
that is, the interface is always flat on average.

%%%%%%%%=======================================================================
\subsection{Second model: spherical constraint}
%%%%%%%%=======================================================================

The next step in the solution of the model consists
of casting the spherical constraint into an equation for $Z(t)$.
To do so, the constraint (\ref{Arc2_contrainte}) is rewritten in Fourier space
\BEQ
\left\langle\!\!\!\left\langle \left\langle \sum_{n=1}^{N}
\left( a_n(t) + b_n(t) \right)^2 \right\rangle \right\rangle\!\!\!\right\rangle
= \frac{1}{N} \left\langle\!\!\!\left\langle \left\langle \sum_{k=0}^{N-1}
\left[ \wht{a}(t,k)\wht{a}(t,k) + \wht{b}(t,k)\wht{b}(t,k) \right]
\right\rangle \right\rangle\!\!\!\right\rangle = N
\EEQ
Initial conditions must be such that the spherical constraint is respected at $t=0$, hence
\BEQ
\frac{1}{N} \left\langle\!\!\!\left\langle  \sum_{k=0}^{N-1} \left[ \wht{a}(0,k)\wht{a}(0,k) + \wht{b}(0,k)\wht{b}(0,k) \right]
\right\rangle\!\!\!\right\rangle
= \frac{1}{2N} \left\langle\!\!\!\left\langle \sum_{k=0}^{N-1} f_{+,00}^2(k) + f_{-,00}^2(k) \right\rangle\!\!\!\right\rangle = N
\EEQ
where the solution (\ref{3.6},\ref{3.7}) was used. From the initial conditions
(\ref{Arc2_initial}) of initially uncorrelated slopes, we have
\BEQ
\left\langle\!\!\left\langle \wht{f}_{\pm,00}(k) \right\rangle\!\!\right\rangle = 0 \;\; , \;\;
\left\langle\!\!\left\langle \wht{f}_{\pm,00}(k)^2 \right\rangle\!\!\right\rangle = N  \;\; , \;\;
\left\langle\!\!\left\langle \wht{f}_{+,00}(k)\wht{f}_{-,00}(k') \right\rangle\!\!\right\rangle =0
\EEQ
The non-vanishing noise correlators read in Fourier space
\BEQ
\left\langle \wht{\eta}^{\pm}(t,k)\wht{\eta}^{\pm}(t',k')\right\rangle
= N \nu T \delta(t-t') \left[ \delta_{k+k',0} \pm \delta_{k-k',0} \right]
\EEQ
such that the constraint can be reexpressed as follows, for this kind of initial condition
\BEA
1 &=& \frac{1}{2N} \sum_{k=0}^{N-1}\left\{ \left[ e^{2\lambda(k) Z(t)}
+ e^{-2\lambda(k) \stackrel{~}{Z}(t)}\right] e^{-4\nu\omega(k)t} \right.
\nonumber \\
& &  \left. +2\nu T \lambda^2(k) \int_0^t \!\D\tau\, e^{-4\nu\omega(k)(t-\tau)}
\left[ e^{2\lambda(k) (Z(t)-Z(\tau))} + e^{-2\lambda(k) (\stackrel{~}{Z}(t)-Z(\tau))}\right] \right\}
\EEA
The asymptotic analysis of this equation is greatly simplified in the continuum limit, when it takes the form
\BEQ \label{3.13}
\int_{0}^{\pi}\!\D k\: \left[ \cosh(2\lambda(k) Z(t))\,e^{-4\nu\omega(k)t} +2\nu T \lambda^2(k) \int_0^t \!\D\tau\:
\cosh(2\lambda(k) (Z(t)-Z(\tau)))\, e^{-4\nu\omega(k)(t-\tau)} \right] =\pi
\EEQ
where the auxiliary functions (\ref{eq:abbrevs}) now stand for their continuum versions
$\omega(k) = 1-\cos k$ and $\lambda(k)=\sin  k$.

In what follows, we shall require the following identities, with $a\in\mathbb{N}$
\BEA
\mathscr{J}_{2a}(A,Z) &:=& \frac{1}{\pi} \int_{0}^{\pi} \!\D k\: e^{A\cos k} \cosh( Z \sin k) \left( \sin k\right)^{2a} =
\frac{\partial^{2a} \mathscr{J}_0(A,Z)}{\partial Z^{2a}}  = \frac{\partial^{2a}  I_0\left( \sqrt{A^2 + Z^2\,}\,\right)}{\partial Z^{2a}}
\nonumber \\
\mathscr{J}_{2a+1}(A,Z) &:=& \frac{1}{\pi} \int_{0}^{\pi} \!\D k\: e^{A\cos k} \sinh( Z \sin k) \left( \sin k\right)^{2a+1} =
\frac{\partial^{2a+1}  I_0\left( \sqrt{A^2 + Z^2\,}\,\right)}{\partial Z^{2a+1}}
\EEA
which are proven in appendix~D and where $I_0$ is a modified Bessel function \cite{Abra65}.
The constraint (\ref{3.13}) can be written more compactly as follows
\BEQ \label{3.15}
e^{-4\nu t} \mathscr{J}_0(4\nu t,2 Z(t)) +2\nu T \int_0^t \!\D\tau\: e^{-4\nu(t-\tau)} \mathscr{J}_2(4\nu(t-\tau),2 Z(t)-2 Z(\tau)) =1
\EEQ
In contrast to the first Arcetri model \cite{Henk15}, or well-known kinetic spherical models \cite{Ronc78,Cugl95,Godr00b},
this equation does not take the form of an easily-solved Volterra integral equation.

%%%%%%%%=======================================================================
\subsection{Second model: observables}
%%%%%%%%=======================================================================
The observables we are interested in are the two-time correlation
and response functions and shall be defined carefully.
%We follow the notations used for the {\sc Arcetri 1u} and {\sc Arcetri 1h} models, in section~1.

For the correlation function, as the order parameter is the local slope $u_n(t)$,
one might expect that $\langle u_n(t)u_m(s)\rangle$ should describe the two-time temporal-spatial correlator
$C_{n,m}(t,s)$. However, a physically sensible definition of correlators must obey two symmetry conditions:
first, for equal times, the purely spatial correlator
$C_{n,m}(t,t)=C_{m,n}(t,t)$ and second, the two-time autocorrelator $C_{n,n}(t,s)=C_{n,n}(s,t)$
are both symmetric. Therefore, we recall the decomposition
$u_n(t) = a_n(t) + b_n(t)$ into an even and an odd part and adopt the
definition\footnote{If we were to consider a complex-valued solution $u_n(t)=a_n(t)+\II b_n(t)$ of the
Burgers equation, see section~2, the definition (\ref{defcorr}) would correspond to $\left\langle u_n(t) u_m(s)^*\right\rangle$.}
\bb\label{defcorr}
C_{n,m}(t,s) := \left\langle\!\left\langle\, \left\langle a_n(t)a_m(s) \right\rangle \, \right\rangle\!\right\rangle
+ \left\langle\!\left\langle \left\langle b_n(t)b_m(s) \right\rangle \right\rangle\!\right\rangle
\ee
Now, using (\ref{3.6},\ref{3.7}) together with the cosine and sine transforms, we find
\BEA
C_{n,m}(t,s) &=& \frac{1}{N} \sum_{k=0}^{N-1} \cos\left(\frac{2\pi}{N}k(n-m)\right) \left[e^{-\stackrel{~}{2}\nu\omega(k)(t+s)}
\cosh \left(\stackrel{~}{\lambda}\!(k)(Z(t)+Z(s))\right) \right. \\
\nn
& &\left. + 2\nu T\int_0^{\min(t,s)} \!\!\D \tau \; \lambda^2(k) e^{-2\nu\omega(k)(t+s-2\tau)}
\cosh \left(\stackrel{~}{\lambda}\!(k)(Z(t)+Z(s)-2Z(\tau))\right) \right]
\EEA
which in the continuum limit $N\rightarrow \infty$ becomes, where $n,m$ are still considered as integers
\bb
\lefteqn{C_{n,m}(t,s) =\frac{1}{2\pi} \int_{-\pi}^{\pi} \cos(k(n-m)) \left[e^{-\stackrel{~}{2\nu}(1-\cos(k))(t+s)}
\cosh \left(\stackrel{~}{\sin}(k)(Z(t)+Z(s))\right) \right.} \\
\nn
& &\left. + 2\nu T\int_0^{\min(t,s)} \!\!\D \tau \; \sin^2(k) e^{-2\nu(1-\cos(k))(t+s-2\tau)}
\cosh \left(\stackrel{~}{\sin}(k)(Z(t)+Z(s)-2Z(\tau))\right) \right].
\ee
The requested symmetries, mentioned above, are now obvious. Furthermore, spatial translation-invariance is now manifest
and we can write $C_{n,m}(t,s) = C_{n-m}(t,s)$.
A more explicit form is obtained by using the identity, valid for $n\in\mathbb{N}$ and $A,Z\in\mathbb{C}$
\BEQ \label{3.19}
\mathscr{C}_n(A,Z) := \frac{1}{\pi}\int_{0}^{\pi} \!\D k\: e^{A\cos k}  \cosh(Z \sin k) \cos(n k) = I_n\left( \sqrt{A^2 + Z^2}\,\right)
\cos \left( n \arctan \left(\frac{Z}{A}\right)\right)
\EEQ
which is proven in appendix~E. This gives, where for notational simplicity we let $t>s$
\bb
C_n(t,s) &=& e^{-2\nu(t+s)}\mathscr{C}_n\left(2\nu (t+s),Z(t)+{Z}(s)\right) \\ \nn
& & + 2\nu T \int_0^s \!\D\tau\: e^{-2\nu(t+s-2\tau)}
\frac{\partial^2}{\partial Z(t)\partial Z(s)}\mathscr{C}_n\left(2\nu(t+s-2\tau), {Z}(t) +Z(s) -2{Z}(\tau)\right)
\ee
In this work, we shall concentrate on the case $T=0$. Then, with (\ref{3.19}), the two-time slope-slope correlator reads explicitly,
in terms of the integrated Lagrange multiplier $Z(t)$
\bb\label{CorrA}
C_n(t,s) &=& e^{-2\nu(t+s)}\mathscr{C}_n\left(2\nu (t+s),Z(t)+Z(s)\right) \\ \nn
&=& e^{-2\nu(t+s)}I_n\left( \sqrt{(2\nu (t+s))^2 + (Z(t)+Z(s))^2}\,\right)
\cos \left( n \arctan \left(\frac{Z(t)+Z(s)}{2\nu (t+s)}\right)\right)
\ee
and we shall extract its long-time scaling behaviour in the next section, after having found $Z(t)$ from (\ref{3.15}). Remarkably,
in the large-time limit $t,s\to\infty$, we shall see that the time-space behaviour simplifies
in the sense that the leading term of the correlator
\BEQ \label{CorrAinf}
C_n(t,s) \simeq C(t,s) \exp\left(-\frac{n^2}{4\nu (t+s)}\right) \cos \left( n \arctan \left(\frac{Z(t)+Z(s)}{2\nu (t+s)}\right)\right)
\EEQ
factorises into the autocorrelator $C(t,s)=C_0(t,s)$ and a second factor, which alone determines the spatial behaviour.

In order to define the linear response of the order-parameter, here identified with the local slope,
a choice of the external perturbation must be made.
Here, we consider the effect of a small perturbation $j_n(t)$, of the slope, on the slope itself.
In generalising the equations of motion, the external
perturbation must be decomposed in the even and odd parts $j_n^{\pm}(t)$, respectively
\BEA
\partial_t a_n(t) &=& \nu \left( a_{n+1}(t) + a_{n-1}(t) -2 a_n(t) \right)
+ \frac{\mathfrak{z}(t)}{2} \left( b_{n+1}(t) - b_{n-1}(t) \right)
\nonumber \\
& & + \demi \left( \eta_{n+1}^{-}(t) - \eta_{n-1}^{-}(t) \right) +j_n^+(t) \\
\partial_t b_n(t) &=& \nu \left( b_{n+1}(t) + b_{n-1}(t) -2 b_n(t) \right)
-  \frac{\mathfrak{z}(t)}{2} \left( a_{n+1}(t) - a_{n-1}(t) \right)
\nonumber \\
& &
-  \demi \left( \eta_{n+1}^{+}(t) - \eta_{n-1}^{+}(t) \right) + j_n^-(t)
%- \textcolor{red}{\eps} \demi \left( \eta_{n+1}^{+}(t) - \eta_{n-1}^{+}(t) \right) + j_n^-(t) % vieille forme, on a eps=+1
\EEA
where $j_n^{+}(t)$ and $j_n^{-}(t)$, respectively, are the conjugate fields associated with the even and odd parts
of the order-parameter $a_n(t)$ and $b_n(t)$. The solution of these equations follows
the same lines which have led to (\ref{3.6},\ref{3.7}), with
the replacements $\lambda(k)\wht{\eta}^{\,\pm} \mapsto \lambda(k)\wht{\eta}^{\,\pm}+ \wht{\jmath}^{\,\mp}$.

The response function is defined as
\BEQ \label{defresp}
R_{n,m}(t,s) := \left\langle\!\!\!\left\langle\left.\frac{\delta a_n(t)}{\delta j_m^+(s)}\right|_{j=0}\right\rangle\!\!\!\right\rangle
+ \left\langle\!\!\!\left\langle \left. \frac{\delta b_n(t)}{\delta j_m^-(s)} \right|_{j=0} \right\rangle\!\!\!\right\rangle
\EEQ
and clearly, only the average over the initial condition (\ref{Arc2_initial}) needs to be taken,
the thermal average becoming trivial. This also implies that the temperature $T$ does not enter
explicitly into the response function.
Inserting the explicit solution, we readily find, also writing the causality condition $t>s$
through the Heaviside function $\Theta$
\BEA
R_{n,m}(t,s) &=& \frac{\Theta(t-s)}{N} \int_0^{t} \!\D\tau \sum_{k,\ell=0}^{N-1} e^{2\nu\omega(k) (t-\tau)}
\cosh\left( \stackrel{\stackrel{~}{}}{\lambda}(k) \left( Z(t)-Z(\tau)\right)\right)
\nonumber \\
& & \times \left[ \cos\left(\frac{2\pi}{N}kn\right)\cos\left(\frac{2\pi}{N}k\ell\right)
+ \sin\left(\frac{2\pi}{N}kn\right)\sin\left(\frac{2\pi}{N}k\ell\right) \right] \delta(\tau-s) \delta_{\ell,m}
\nonumber \\
&=& \frac{\Theta(t-s)}{N} \sum_{k=0}^{N-1} e^{2\nu\omega(k) (t-s)}
\cosh\left( \stackrel{\stackrel{~}{}}{\lambda}(k) \left( Z(t)-Z(s)\right)\right)
\cos\left(\frac{2\pi}{N}k(n-m)\right) ~~~~
\EEA
which is spatially translation-invariant, as it should be, hence $R_{n,m}(t,s)=R_{n-m}(t,s)$.
In the $N\to\infty$ limit, this simplifies into, using (\ref{3.19})
\BEA \label{RespQ}
\lefteqn{ R_{n}(t,s) = \frac{\Theta(t-s)}{\pi}\int_0^{\pi}\!\D k\: e^{-2\nu(1-\cos k) (t-s)} \cosh\left( \left(Z(t)-Z(s)\right)
\stackrel{\stackrel{~}{}}{\sin} k\right) \cos k n} \\
&=& \Theta(t-s)\: e^{-2\nu(t-s)} I_{n}\left( \sqrt{ 4\nu^2(t-s)^2 +\left( Z(t)-Z(s)\right)^2\,}\,\right)
\cos\left( n \arctan\left(\frac{Z(t)-Z(s)}{2\nu(t-s)}\right) \right)~~~  \nonumber
\EEA
In the next section, using (\ref{3.15}), the asymptotic long-time scaling behaviour will be analysed.
Again, we find that for large times, the leading term simplifies
\BEQ\label{RespQinf}
R_n(t,s) \simeq R(t,s) \exp\left(-\frac{n^2}{4\nu (t-s)}\right) \cos\left( n \arctan\left(\frac{Z(t)-Z(s)}{2\nu(t-s)}\right) \right)
\EEQ
into a product of the autoresponse $R(t,s)=R_0(t,s)$ and a second factor, which determines the spatial dependence alone.

Alternatively, one might also consider how the local slope will respond to a
small change in the height variable. In this case, it is enough
to replace $\eta\mapsto \eta+j$ in the equations of motion (\ref{Arcetri2}).
The formal definition of the response function is still given
by (\ref{defresp}), although the physical meaning of the external fields $j_n^{\pm}(t)$
has changed. The explicit calculation is analogous
to the previous ones and we just quote the result
\BEA
\lefteqn{{\cal R}_{n}(t,s) = \frac{\Theta(t-s)}{N}\sum_{k=0}^{N-1} \lambda(k)\, e^{-2\nu\omega(k)(t-s)}
\sinh\left( \stackrel{\stackrel{~}{}}\lambda(k)
\left( Z(t)-Z(s)\right) \right) \cos\left(\frac{2\pi}{N} kn\right) }
\nonumber \\
&\stackrel{N\to\infty}{=}& \frac{\Theta(t-s)}{\pi}\int_0^{\pi} \!\D k\: e^{-2\nu(1-\cos k)(t-s)}
\sinh\left( \left( Z(t)-Z(s)\right) \stackrel{\stackrel{~}{}}{\sin} k \right) \sin (k) \cos (kn)
\label{3.29} \\
&=& \Theta(t-s)\: e^{-2\nu(t-s)} \times \nonumber \\
& & \times \frac{\partial}{\partial Z(t)} \left[I_{n}\left( \sqrt{ 4\nu^2(t-s)^2 +\left( Z(t)-Z(s)\right)^2\,}\,\right)
\cos\left( n \arctan\left(\frac{Z(t)-Z(s)}{2\nu(t-s)}\right) \right)\right]~~~  \nonumber
\EEA
where we have used (\ref{3.19}) again. The asymptotic behaviour follows from the one of $Z(t)$.

%%%%%%%%=======================================================================
\subsection{Third model}
%%%%%%%%=======================================================================

In order to define the observables of the third Arcetri model, re-write first the
(anti-)symmetrised equations of motion in Fourier space
\BEA
\partial_t \wht{a}(t,k) &=& -2\nu \omega(k) \wht{a}(t,k)
+ \mathfrak{z}(t)\lambda(k) \wht{b}(t,k)
+ \wht{\eta}^{+}(t,k)
\nonumber \\
\partial_t \wht{b}(t,k) &=& -2\nu \omega(k) \wht{b}(t,k)
+ \mathfrak{z}(t)\lambda(k) \wht{a}(t,k)
+  \wht{\eta}^{-}(t,k) ~~~~~~~
\EEA
where we used the abbreviations (\ref{eq:abbrevs}).  %and we have also added a small perturbation $\wht{\jmath}^{\,\pm}$ of the height.
The solution of these equations reads
\BEA
\left. \vekz{\wht{a}(t,k)}{\wht{b}(t,k)}\right\} &=& \demi \left[ \wht{f}_{+,00}(k)\: e^{-2\nu\omega(k) t+\lambda(k) Z(t)}
\pm \wht{f}_{-,00}(k)\: e^{-2\nu\omega(k) t-\lambda(k) Z(t)} \right]
\nonumber \\
& & +\demi \int_0^t \!\D\tau\,  \left[ \left( \wht{\eta}^{+}(\tau,k)+\wht{\eta}^{-}(\tau,k)\right)
e^{-2\nu\omega(k)(t-\tau) + \lambda(k) \left( Z(t) - Z(\tau)\right)} \right. \nonumber \\
& & \left. \hspace{2.truecm}\pm \left( \wht{\eta}^{+}(\tau,k)-\wht{\eta}^{-}(\tau,k)\right)
e^{-2\nu\omega(k)(t-\tau) - \lambda(k) \left( Z(t) - Z(\tau)\right)} \right]
\label{Arc3eqmouv}
\EEA
where the upper signs correspond to $\wht{a}$ and the lower signs to $\wht{b}$. The spherical constraint is now give by eq.~(\ref{Arc3_contrainte})
and takes the form
\BEQ
\frac{1}{N}
\left\langle\!\!\!\left\langle \left\langle \sum_{k=0}^{N-1}
\left[ \lambda(k)^2 \left[ \wht{a}(t,k)\wht{a}(t,k) + \wht{b}(t,k)\wht{b}(t,k)\right] \right]
\right\rangle \right\rangle\!\!\!\right\rangle = N
\EEQ
This can be evaluated along the same lines as before. We merely quote the end result, in the continuum limit
\BEQ \label{3.33}
e^{-4\nu t} \mathscr{J}_2(4\nu t,2 Z(t)) +2\nu T \int_0^t \!\D\tau\: e^{-4\nu(t-\tau)} \mathscr{J}_2(4\nu(t-\tau),2 Z(t)-2 Z(\tau)) =1
\EEQ

If $T=0$, the conservation laws (\ref{2.8}) obtained for the second model also hold for the third.
This implies a constant height profile $\langle\!\langle h(t,r)\rangle\!\rangle=H_0$, where for simplicity,
we used the initial conditions (\ref{Arc3_initial}). From now on, we set $H_0=0$ in (\ref{Arc3_initial}), without restriction to the generality.

The time-space correlator is defined analogously as in the second model, by eq.~(\ref{defcorr}),
but now using the decomposition $h_n(t)=a_n(t)+b_n(t)$ of the height
in an even and an odd part. Although we re-use the formal definition eq. (\ref{defcorr}),
the physical interpretation is now a different one and gives a height-height correlator.
The main computational difference with respect to the second model is the
absence of the factor $\lambda(k)$ before the thermal noise $\wht{\eta}^{\pm}$.
For the initial conditions (\ref{Arc3_initial}), spatial translation-invariance holds for all times $t,s>0$, so that we can write
$C_{n-m}(t,s)=C_{n,m}(t,s)$. We finally have, using (\ref{3.15})
\BEA \label{3.34}
C_n(t,s) &=& e^{-2\nu(t+s)}\mathscr{C}_n(2\nu (t+s),Z(t)+Z(s)) \\ \nn
& & + 2\nu T \int_0^{\min (t,s)} \!\D\tau\: e^{-2\nu(t+s-2\tau)}
\mathscr{C}_n(2\nu(t+s-2\tau), Z(t) +Z(s) -2 Z(\tau))
\EEA
In particular, for $T=0$, we recover the same abstract expression, eq.~(\ref{CorrA}), as for the slope-slope correlator $C_n^{(II)}(t,s)$
but with the difference that $Z(t)$ now has to be found from the spherical constraint (\ref{3.33}).

For the calculation of the linear response of the height with respect to a small change in the height,
one replaces $\wht{\eta}^{\pm} \mapsto \wht{\eta}^{\pm} + \wht{\jmath}^{\,\pm}$ in the equation of motion (\ref{Arc3eqmouv}).
We can simply re-use the definition (\ref{defresp}), and formally recover the abstract form
\BEA
R_{n}(t,s) &=& \Theta(t-s)\: e^{-2\nu(t-s)} \times \nonumber \\
& & \times I_{n}\left( \sqrt{ 4\nu^2(t-s)^2 +\left( Z(t)-Z(s)\right)^2\,}\,\right)
\cos\left( n \arctan\left(\frac{Z(t)-Z(s)}{2\nu(t-s)}\right) \right)~~~~~
\label{3.35}
\EEA
identical to (\ref{RespQ}).
In particular, in the long-times limit and for $T=0$, both the height correlator and the height response factorise, as
in (\ref{CorrAinf},\ref{RespQinf}), respectively. Again, $Z(t)$ is found from (\ref{3.33}).

%%%%%%%%%%%%%%%%%%%%%%%%%%%%%%%%%%%%%%%%%%%%%%%%%%%%%%%%%%%%%%%%%%%%%%%%%%%%%%%%
\section{Long-time behaviour}
%%%%%%%%%%%%%%%%%%%%%%%%%%%%%%%%%%%%%%%%%%%%%%%%%%%%%%%%%%%%%%%%%%%%%%%%%%%%%%%%

%%%%%%%%=======================================================================
\subsection{Spherical constraint at $T=0$}
%%%%%%%%=======================================================================

First, we have to determine the long-time behaviour of the Lagrange multiplier $Z(t)$, from the
spherical constraints eqs.~(\ref{3.15},\ref{3.33}). From now on, we shall restrict to the case $T=0$.

For the second model, eq.~(\ref{3.15}) reduces to
\BEQ \label{4.1}
e^{-4\nu t} I_0\left(4\nu t\sqrt{1 + \left(\frac{ Z(t)}{2\nu t}\right)^2}\:\right) =1.
\EEQ
In appendix~B, we shall show that for large times $Z(t) \simeq \sqrt{\nu t \ln(8\pi\nu t)\,}$.

For the third model, we must solve eq.~(\ref{3.33}). In appendix~B it is shown that this is equivalent to
\BEA \label{4.2}
& & \left[ I_1\left(\sqrt{(4\nu t)^2 + (2 Z(t))^2}\:\right)(4\nu t)^2 \right. \nonumber \\
& & \left.+ 
2 Z(t)^2\sqrt{ (4\nu t)^2 + (2 Z(t))^2\,}\, \left( I_0\left(\sqrt{(4\nu t)^2 + (2 Z(t))^2}\:\right)+I_2\left(
\sqrt{(4\nu t)^2 + (2 Z(t))^2}\:\right)\right)\right] \nonumber \\
&=& e^{4\nu t} {\left[ (4\nu t)^2 + (2Z(t))^2\right]^{-3/2}} 
\EEA
To leading order, this would give the solution $Z(t) \simeq \sqrt{3\nu t \ln((32\pi e)^{1/3}\: \nu t)\,}$. 
Then, this could be combined with the second model as follows:
for $t$ large enough, we have
\BEQ \label{4.3}
Z(t) = \sqrt{ \kappa_1 \nu t\: \ln(\kappa_2 t)\,} \;\; , \;\;
\left\{ \begin{array}{lll}
\kappa_1 = 1\;\; , \;\; & \kappa_2 = 8\pi\nu & \mbox{\rm ~second model} \\
\kappa_1 = 3\;\; , \;\; & \kappa_2 = (32\pi e)^{1/3}\nu & \mbox{\rm ~third model}
\end{array} \right.
\EEQ
However, as also shown in appendix~B, it is advisable to include the next-to-leading terms as well. 
If that is done, we find, for the third model and $t$ large enough
\BEQ \label{4.4}
Z(t) \simeq \sqrt{2t W\left( \left(32\pi e\right)^{1/2} t^{3/2}\right) -t\,} 
\simeq \sqrt{3t\ln (\kappa_2 t) \left[ 1 - \frac{2}{3}\frac{\ln (\frac{3}{2} \ln \kappa_2 t)}{\ln \kappa_2 t} -\frac{1}{3}\frac{1}{\ln \kappa_2 t} \right]\,}
\EEQ
where $W(x)$ denotes Lambert's $W$-function \cite{Lambert1758,Corl96}. 
The results (\ref{4.3}), and (\ref{4.4}) for the third model, are the basis of the entire asymptotic analysis.
Formally, for truly enormous times $t\ggg 1$, the distinction between the second and the third model merely comes from the
values of the two constants $\kappa_{1,2}$, as given in (\ref{4.3}), and both models can be analysed together.
However, the further logarithmic corrections in (\ref{4.4}) will lead to some significant differences between the second and the third model, even for $T=0$, 
as we shall now see. 
In the main text, we merely quote our results and refer to appendices~B and~C for the calculations.

%%%%%%%%=======================================================================
\subsection{Zero-temperature correlators}
%%%%%%%%=======================================================================

We now turn to the correlators.
For a vanishing temperature $T=0$, we have already shown in (\ref{CorrA},\ref{CorrAinf}) that the time-space correlator $C_n(t,s)$ factorises
into the  autocorrelator $C(t,s)$ and a space-dependent part. The autocorrelator takes the form
\BEQ
C(t,s) = e^{-2\nu(t+s)}\:
{I_0\left( 2\nu(t+s)\sqrt{1+\left(\frac{Z(t)+Z(s)}{2\nu(t+s)}\right)^2\,}\,\right)}
\EEQ
where $Z(t)$ is given by (\ref{4.3}). This autocorrelator does {\em not}\/ obey the scaling of simple ageing,
where $t,s\to\infty$ and $y=t/s>1$ is kept fixed. We rather must consider a different scaling behaviour,
where again $t,s\to\infty$ such that a certain scaling variable $y$ is being kept fixed. We find two possibilities:
\begin{enumerate}
\item the time difference is given by\footnote{The notation is chosen such that one returns to simple ageing, if the logarithmic factors are dropped.}
\BEQ \label{Ascaling}
\tau = t-s = (y-1)\, \frac{s}{\ln^{1/2} \kappa_2 s}.
\EEQ
and we have the scaling form
\BEQ \label{Afonction}
C(t,s) = C^{(0)} \left( \ln s\right)^{(\kappa_1-1-\delta_{\kappa_1,3})/2} \exp\left( - \frac{\kappa_1}{32}(y-1)^2 \right)
\EEQ
with the constant $C^{(0)}=\sqrt{ \kappa_2^{\kappa_1}/(8\pi\nu)\,}$ for the second model and
$C^{(0)}=\sqrt{ \kappa_2^{\kappa_1}e^{1/2}/(12\pi\nu)\,}\,$ for the third model. The scaling (\ref{Ascaling},\ref{Afonction}))
was seen before in the phase-separation kinetics of the spherical model,
at temperature $T=0$, with a conserved order parameter (model B dynamics) \cite{Bert00}.
\item the time difference is given by, with $\vartheta>\demi$
\BEQ \label{Cscaling}
\tau = t-s = \frac{s}{\ln^{1/2}(\kappa_2 s)}
\sqrt{ W\left( (y-1)^2\, \ln^{1-2\vartheta}(\kappa_2 s) \right)\:}
\EEQ
where $W(x)$ is, again, Lambert's $W$-function. For $s\to\infty$, this gives $\tau\simeq (y-1) s \ln^{-\vartheta}(\kappa_2 s)$.
The autocorrelator becomes
\BEQ \label{Cfonction}
C(t,s) = C^{(0)} \left( \ln(\kappa_2 s)\right)^{(\kappa_1-1-\delta_{\kappa_1,3})/2}
\EEQ
As we shall see, we must further distinguish here the cases $\demi<\vartheta<1$ and $\vartheta>1$.
\end{enumerate}
A scaling behaviour according to (\ref{Ascaling},\ref{Cscaling}) corresponds to {\em logarithmic sub-ageing}, since the time difference
$\tau=t-s$ grows more slowly than $s$, by a logarithmic factor.\footnote{Simple inequalities
exclude the opposite case of `super-ageing' where $t-s$ would grow faster than $s$ \cite{Kurc02}.}
Although the forms (\ref{Ascaling},\ref{Cscaling}) are distinct from simple ageing as described in section~1, 
we shall cast the autocorrelator into a scaling form
$C(t,s)=s^{-b}\ln^{-\hat{b}}s\; f_C(y)$ and shall check for an asymptotic form 
$f_C(y)\sim y^{-\lambda_C/z} \ln^{\hat{\lambda}_C/z} s$ as $y\to\infty$. If these
forms apply, the exponents $b$, $\lambda_C$ are quoted in tables~\ref{tab1} and~\ref{tab2}. By analogy with equilibrium critical phenomena,
we also introduce the {\em logarithmic sub-scaling exponents $\hat{b}$, $\hat{\lambda}_C$} \cite{Kenn06}. 
We find $\hat{b}=\hat{\lambda}_C=0$ for the second Arcetri model and $\hat{b}=-\demi$, $\hat{\lambda}_C=0$ for the third Arcetri model.

Concerning the equal-time autocorrelator $C(t,t)=\left(\ln(\kappa_2 t)\right)^{(\kappa_1-1-\delta_{\kappa_1,3})/2}$ 
in both scaling regimes, there is a difference in
interpretation between the second and the third model. In the second model, $C(t,t)=1$ because of the constraint (\ref{4.1}), which is consistent
with a probabilistic interpretation, either in terms of the slopes or else in terms of particles and holes. Indeed, one has $\kappa_1=1$ in this case.
For the third model, $C(t,t)=w^2(t)=\left( \ln \kappa_2 t\right)^{1/2}$ is simply the square of the interface width
$w(t)\sim t^{\beta}\ln^{\hat{\beta}}t$. Hence
for $d=1$, where $\kappa_1=3$, the interface is logarithmically rough, with growth exponents $\beta=0$ and $\hat{\beta}=\frac{1}{4}$.

%%++++++++++++++++++++++++++++++++++++++++++++++++++++++++++++++++++++++++++++++++
\begin{figure}[tb]
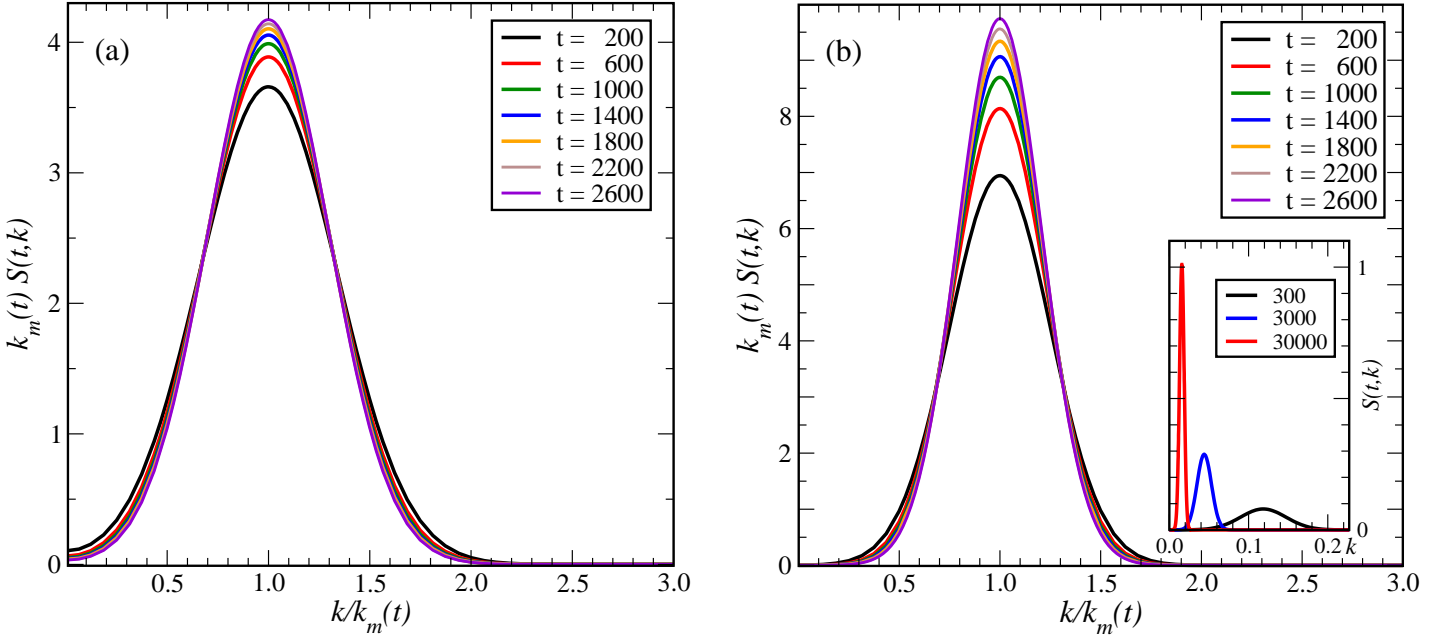

\centerline{\psfig{figure=durang6_fig4_FS_Arc2_U,width=3.6in,clip=}~~~~\psfig{figure=durang6_fig4_FS_Arc3_H,width=3.6in,clip=}}
\caption[fig4]{Scaled structure factor $S(t,k)$, normalised to $S(1,0)=1$, for (a) the slope-slope correlations in the second model and (b)
the height-height correlations in the third model, for several times $t$ and at temperature $T=0$.
In the inset in (b) also shows the unscaled structure factor (arbitrary units), for several times $t$.
\label{fig4}
}
\end{figure}
%%++++++++++++++++++++++++++++++++++++++++++++++++++++++++++++++++++++++++++++++++

The time-space-dependent slope-slope correlator $C_n(t,s)$ is given by (\ref{CorrAinf}).
This yields the following long-time behaviour
\BEQ \label{ACorrel2}
C_n(t,s) \simeq C(t,s) \exp\left(-\frac{1}{2} \left(\frac{n}{\sqrt{2\nu(t+s)\,}\,}\right)^2\right)
\cos\left( \frac{n}{\sqrt{2\nu(t+s)}} \sqrt{\frac{\kappa_1}{2}(\ln \kappa_2 t + \ln \kappa_2 s)\,}\right)
\EEQ
In particular,  this gives the equal-time correlator
\BEA
C_n(t) &:=& C_n(t,t) =
e^{-4\nu t} {I_n\left( 4\nu t\sqrt{1+\left(\frac{Z(t)}{2\nu t}\right)^2\,}\,\right)\cos\left(n\arctan \frac{Z(t)}{2\nu t}\right)}
\nonumber \\
&\stackrel{t\to\infty}{\simeq} & C_0(t,t) \exp\left( - \left(\frac{n}{L_1(t)}\right)^2\right)\, \cos\left( \frac{n}{L_2(t)}\right)
\label{ACorrel1}
\EEA
with the equal-time autocorrelator $C_0(t) := C_0(t,t)$ and the two distinct length scales
\BEQ
L_1(t) := \sqrt{8\nu t\,} \;\; , \;\; L_2(t) := \sqrt{ \frac{4\nu}{\kappa_1} \frac{t}{\ln (\kappa_2 t) \digamma(t)}\,}
\EEQ
where $\digamma(t)$ is defined in appendix~B and gives double-logarithmic corrections to scaling for the third model. The presence of two logarithmically different
length scales indicates a  breaking of dynamical scaling. This becomes even more explicit when considering the structure factor 
\BEA
S(t,k) &=& \frac{1}{\sqrt{2\pi\,}\,}\int_{\mathbb{R}} \!\D n\: e^{-\II k n}\, C_n(t) 
\nonumber \\
&=& \frac{C_0(t) L_1(t)}{\sqrt{2}} \exp\left(-\frac{1}{4}\frac{L_1^2(t)}{L_2^2(t)}\right)\exp\left(-\frac{L_1^2(t)}{4} k^2\right)
\cosh\left( \frac{L_1(t)}{L_2(t)}\frac{L_1(t) k}{2}\right) 
\label{eq:FS} \\
&\simeq& \frac{C_0(t) L_1(t)}{\sqrt{2\,}\,(\kappa_2 t)^{\kappa_1/2}} e^{-\left(L_1(t) k/2\right)^2} \cosh\left( \frac{L_1(t)}{L_2(t)}\frac{L_1(t) k}{2}\right) 
\left\{ \begin{array}{ll} 1 & \mbox{\rm ~~;~ second model} \\
                          \frac{3e^{1/2}}{2} \ln\kappa_2 t & \mbox{\rm ~~;~ third model}
        \end{array} \right.
\nonumber
\EEA 
where the last logarithmic factor comes  from the auxiliary function $\digamma(t)$. 
Working out the long-time behaviour of the two lengths, we find, for the second and third model, respectively 
\begin{subequations} \label{4.11}
\begin{align}
S^{(II)}(t,k) &= S^{(0)}\: {e^{-2\nu t k^2}}\, \cosh\left(\sqrt{4\nu t\ln 8\pi\nu t\:}\: k\right) \\
S^{(III)}(t,k) &= S^{(0)}\: e^{-2\nu t k^2}\, \left(\frac{\ln^{3/2} \kappa_2 t}{t}\right) 
\cosh\left(k\sqrt{4\nu t\left( 3\ln \kappa_2 t-2\ln(\ln\kappa_2 t) -\left(1+2\ln\frac{3}{2}\right)\right)\:}\: \right) 
\end{align}
\end{subequations}
with known constants $S^{(0)}$ and $\kappa_2$ was defined above. 

The explicit expressions (\ref{4.11}) permit a clear understanding of the distinct length scales involved. Since for both models, the structure factor
contains two factors with different $k$-dependence, one expects a peak at some time-dependent {\it lieu} $k_m(t)$. 
Figure~\ref{fig4} shows that this indeed the case, notably
in the inset of figure~\ref{fig4}b, which illustrates how for increasing times the peak becomes sharper and is progressively shifted towards $k\to 0$.
Eq.~(\ref{eq:FS}) implies that $k_m(t) \simeq L_2^{-1}(t)\sim \left(\frac{\ln t}{t}\right)^{1/2}$. 
If one attempts to scale the structure factors with respect to $L_2(t)$, as is done 
in figure~\ref{fig4}ab, one might believe at first sight that a scaling behaviour would result (at least for the second model). 
However, the presence of the diffusive length $L_1(t)\sim t^{1/2}$, it is impossible to achieve a collapse and 
dynamical scaling does {\em not} hold for {\em all} wave numbers $k\geq 0$.\footnote{In the third model, 
scaling is further broken by an additional logarithmic factor.} 
This kind of behaviour is completely analogous to the one of phase-separation in the
$T=0$ kinetic spherical model with a conserved order-parameter (model B dynamics) \cite{Coni94,Bray92}.
These two length scales also describe the time-space correlator: while $L_{2}$ is seen in the spatial modulation, the scale $L_{1}$ describes
the overall spatial decay.

The scaling of the autocorrelator introduces at least one more scale $L_{\rm corr}(t)\sim \left(t/\ln^{\vartheta} t\right)^{1/2}$,
with $\vartheta\geq\demi$. Several distinct regimes must be distinguished, as we shall see in section~5.

%%%%%%%%=======================================================================
\subsection{Zero-temperature response}
%%%%%%%%=======================================================================

{}From (\ref{RespQ},\ref{RespQinf}), we have for large times the factorisation into the autoresponse
\BEQ
R(t,s) = e^{-2\nu(t-s)}
{I_0\left( 2\nu(t-s)\sqrt{1+\left(\frac{Z(t)-Z(s)}{2\nu(t-s)}\right)^2\,}\,\right)}
\EEQ
and the time-space response
\BEA \label{QCorrel2}
R_n(t,s) &=& e^{-2\nu(t-s)}
{I_n\left( 2\nu(t-s)\sqrt{1+\left(\frac{Z(t)-Z(s)}{2\nu(t-s)}\right)^2\,}\,\right)\cos\left(n\arctan \frac{Z(t)-Z(s)}{2\nu(t-s)}\right)}
\nonumber \\
&\simeq & R(t,s) \exp\left[-\demi\left(\frac{n}{\sqrt{2\nu (t-s)\,}\,}\right)^2 \right]
\EEA
The scaling is obtained as follows and corresponds to (\ref{Cscaling}).\footnote{The only difference between the second and third model is a logarithmic
 modification of the scaling variable, as explained in appendix~C, and which disappears asymptotically for times $t,s\ggg 1$.} We have for the time difference
\BEQ \label{Qscaling}
\tau = t-s =  \frac{s}{\ln (\kappa_2 s)}\: W\left( (y-1)\ln^{1-\vartheta} (\kappa_2 s) \right)
\EEQ
where $\vartheta>1$, in terms of Lambert's $W$ function. This gives $\tau\sim (y-1) s\ln^{-\vartheta}(\kappa_2 s)$ for large times.
The autoresponse (\ref{RespQ}) takes the scaling form
\BEQ \label{Qfunction}
R(t,s) = (4\pi\nu s)^{-1/2} \ln^{\vartheta/2}(\kappa_2 s)\; (y-1)^{-1/2}
\EEQ
which is so close to the one found in systems undergoing simple ageing (as the first Arcetri model, see appendix~A), up to a logarithmic prefactor, that
we read off the exponents $a=-\demi$ and $\lambda_R/z=\demi$, see table~\ref{tab1}.
There is no correspondence in the autoresponse to the scaling (\ref{Ascaling}) of the autocorrelator.

We also observe that the response ${\cal R}_n(t,s)$ defined in (\ref{3.29}) has the same factorisation into the autoresponse ${\cal R}(t,s)$ and a spatial part as in
(\ref{RespQinf}). The autoresponse is readily found and reads, in the scaling limit with $\vartheta>1$
\BEQ
{\cal R}(t,s) =\left(8\pi^{1/2}\nu s\right)^{-1}\:
 \ln^{(1+\vartheta)/2}(\kappa_2 s)\: (y-1)^{-1/2}
\EEQ
and we read off $a_{\cal R}=0$ and $\lambda_{\cal R}=1$.

In the time-space responses, we merely find a single further length scale $L_{\rm diff}(t)\sim t^{1/2}$, which describes the overall decay, but no spatial
modulation of the response.

%%%%%%%%%%%%%%%%%%%%%%%%%%%%%%%%%%%%%%%%%%%%%%%%%%%%%%%%%%%%%%%%%%%%%%%%%%%%%%%%
\section{Discussion and perspectives}
%%%%%%%%%%%%%%%%%%%%%%%%%%%%%%%%%%%%%%%%%%%%%%%%%%%%%%%%%%%%%%%%%%%%%%%%%%%%%%%%

Triggered by an analogy with the spherical model of a ferromagnet \cite{Berl52}, we have used the Burgers and Kardar-Parisi-Zhang equations to
define two new models, which we have called the {\em second Arcetri model} and the {\em third Arcetri model}, see sections 2.2 and 2.3. Because of
the  natural initial conditions (\ref{Arc2_initial}) and (\ref{Arc3_initial}), respectively, the second model is interpreted as a lattice gas model, whereas
the third model appears to describe a growing interface. We have found exactly, at vanishing `temperature' $T=0$, the exact two-time correlators and responses.
Unexpectedly, the scaling behaviour of these turned out not to be given by simple ageing, but rather by a subtle modification of this, described by
{\em logarithmic sub-ageing} and characterised by the presence of several logarithmically distinct time-dependent length scales.

%%++++++++++++++++++++++++++++++++++++++++++++++++++++++++++++++++++++++++++++++++
\begin{figure}[tb]
\centerline{\psfig{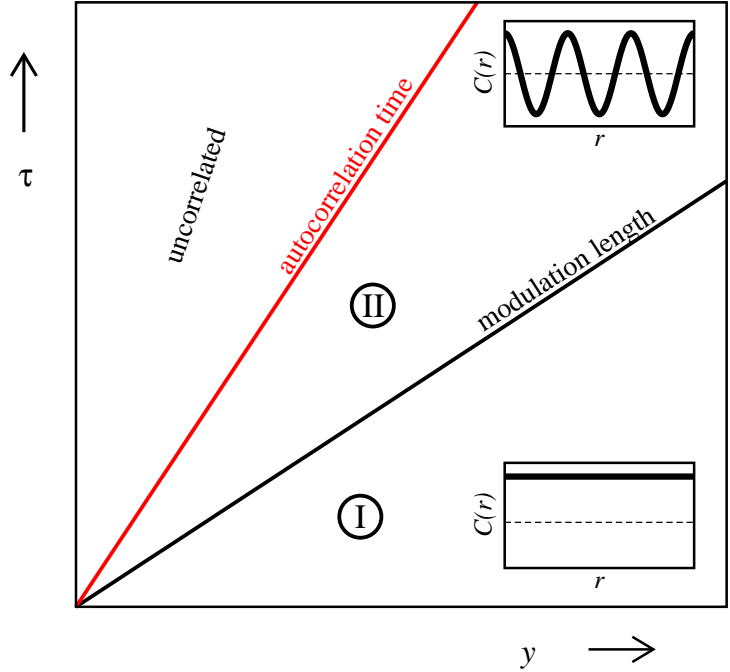}}
\caption[fig3]{Schematic kinetic phase diagram of the second and third Arcetri models at temperature $T=0$. The regions of distinct scaling behaviour
for time differences $\tau=t-s=\tau(s,y)$ are indicated. The spatial behaviour of the single-time correlator in shown in the insets for the phases I and II.
\label{fig3}
}
\end{figure}
%%++++++++++++++++++++++++++++++++++++++++++++++++++++++++++++++++++++++++++++++++

As we shall see now, this is a very fortunate circumstance, since the separation of several length scales, 
which coalesce for simple ageing, allows for a much more
clear understanding how the ageing process takes place. Figure~\ref{fig3} summarises this as a kinetic phase diagram.

\begin{figure}[tb]
\centerline{\psfig{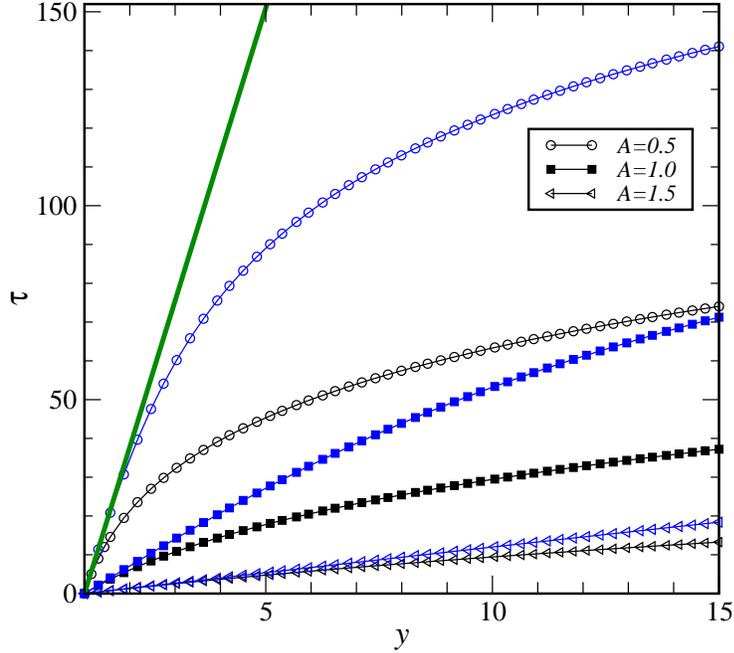}}
\caption[fig2]{Illustration of the definition of the scaling variables $\tau=\tau(s,y)$, for fixed waiting time $s$. The
black curves show the definition (\ref{Qscaling}) for the response, for several values of $\vartheta=2A$, and the blue curves correspond to
(\ref{Cscaling}) for the correlator. The straight green line is the asymptotic form $\tau=(y-1) s/\ln s$.
\label{fig2}
}
\end{figure}

Equations in physics should be read as instructions how to carry out experiments. 
In the case of ageing, the central equation describes the scaling of the difference between {\em observation time} $t$ and {\em waiting time} $s$:
\BEQ \label{5.1}
\tau = t-s = \frac{s}{\ln^{\vartheta} s} f(s,y) \stackrel{s\to\infty}{\simeq} \frac{s}{\ln^{\vartheta} s} (y-1)
\EEQ
in terms of a certain function $f=f(s,y)$ which we have determined explicitly, for the second and third Arcetri models, in section~4.
In combination with the scaling form of the autocorrelator $C=C(t,s)$ or the autoresponse $R=R(t,s)$, the meaning of eq.~(\ref{5.1}) is:
\begin{quote}
{\it ``For a large waiting time $s$, and a fixed scaling variable $y>1$,
compute the observation-time scale $t=s+\tau(s,y)$ in order to observe the corresponding scaling behaviour of the observable in question.''}.
\end{quote}
For systems with logarithmic sub-ageing, as it is realised in the second and third Arcetri models at $T=0$, this leads to the following insights:
\begin{enumerate}
\item Phase I is characterised by an exponent $\vartheta>1$ in (\ref{5.1}).\footnote{The precise scaling prescriptions are given by
(\ref{Cscaling},\ref{Qscaling}), see also figure~\ref{fig2}.} Both correlators and responses scale, and the asymptotic forms of the
scaling variables are compatible. The scaling functions are given by
eqs.~(\ref{Cfonction}) and (\ref{Qfunction}), respectively and the corresponding exponents are listed in tables~\ref{tab1} and~\ref{tab2}.
These relatively small time differences also imply a corresponding length scale
$L_{\rm corr}(t)\sim \sqrt{ t/\ln^{\vartheta}t\,}$. On these time and space scales, the autocorrelation is perfect and the system is
spatially homogeneous. The most rapid events occur in the slow decay of the response to an external, localised perturbation.
\item The end of phase I is seen when one goes to larger scales by choosing $\vartheta=1$. At this length scale, which seen from phase I would correspond to
a $y\to\infty$ limit, the responses have decayed. The new feature is the onset of spatial modulation of the spatial correlators (\ref{ACorrel1}),
which occurs at the scale $L_{\max}\sim \sqrt{t/\ln t\,},$ and signalled by a strong peak in the structure function at wave number $k_{\max}$,
see figure~\ref{fig3}. Since the scale $L_{\rm diff}(t)\gg L_{\max}(t)$, these modulations occur with constant amplitude. And because the scaling
form (\ref{Cscaling}) of the autocorrelator still holds, autocorrelations do not dissipate away.
\item Phase II is characterised by the intermediate range $\demi<\vartheta<1$, between the onset of spatial modulations and a new scaling
form of the autocorrelator, which in this scaling limit is still perfect. In contrast to phase I, the responses have at this time-scale decayed away
and do not scale anymore.
\item The end of phase II is seen at large length scales $L_{\rm corr}(t)$ which correspond to $\vartheta=\demi$.
At these scales, the system consists of many small spatial units, each of them still fully ordered.
Now the autocorrelator decays according to the scaling form (\ref{Ascaling}), but since still $L_{\rm diff}\gg L_{\rm corr}(t)$, each spatial unit
remains fully ordered. Responses to external perturbations are vanishingly small on these scales.
\item On scales so large that they correspond to $0<\vartheta<\demi$, there is no more scaling form and temporal correlations are lost. The eventual decay
of the amplitude of spatial correlations at distances corresponding to $\vartheta=0$ has no effect on a system which at these scales is already disordered.
\end{enumerate}

A completely analogous behaviour can be found in the kinetic spherical model with a conserved order parameter (model B dynamics) at temperature $T=0$, which
describes spinodal decomposition, in $d>2$ dimensions. It has a dynamical exponent $z=4$, hence is distinct from the Arcetri models.
However, the breaking of scale-invariance of the equal-time spin-spin correlator, through two logarithmically distinct length scales, is analogous
to (\ref{ACorrel1}) \cite{Coni94}. Furthermore, the scaling of the magnetic autocorrelator $C(t,s)$ and of the magnetic autoresponse $R(t,s)$ \cite{Bert00}
is completely analogous to the Arcetri models. The scaling variable is, again, defined by $\tau=t-s=(y-1)s\ln^{-\vartheta}s$. If $\vartheta>1$,
for $s$ sufficiently large, we find the autocorrelator $C(t,s)=1$ 
and the autoresponse $R(t,s)\sim s^{-(d+2)/4} \left(\ln^{\vartheta (d+2)/4}s\right) (y-1)^{-(d+2)/4}$.
If $\demi<\vartheta<1$, one still has $C(t,s)\sim\exp\left(-\frac{d}{64}(y-1)^2\ln^{1-2\vartheta}s\right)\stackrel{s\to\infty}{\to} 1$
and the autoresponse does not scale. Finally, for $\vartheta=\demi$, one has
$C(t,s)\sim \exp\left(-\frac{d}{64}(y-1)^2\right)$ \cite{Bert00}.\footnote{The presence of two logarithmically different length scales is called
{\it multiscaling} \cite{Coni94}. In ${\rm O}(n)$-symmetric magnets with a conserved order-parameter, quenched to temperature $T=0$,
multiscaling only occurs exactly in the spherical limit $n=\infty$ but does not arise for $n$ finite \cite{Bray92,Bray94,Maze06}. 
It appears plausible that a similar effect may also arise for the second and third Arcetri models at $T=0$, in view of the conservation law (\ref{2.8}). 
Multiscaling has been argued to be present in systems such as diffusion-limited aggregation ({\sc dla})
\cite{Coni90}, but apparently no consensus has been reached on whether multiscaling in {\sc dla} is genuine \cite{Moha09}
or rather an effective finite-size effect \cite{Somf03,Mens12}. 

A slightly different situation arises in the {\em quantum} dynamics of the quantum spherical model, 
where the associated Lindblad equation preserves the canonical commutator relations, which might be viewed as an (infinite) set of prescribed
conservation laws. Then, at $T=0$ and for quantum quenches deep into the two-phase coexistence region, a simple scaling behaviour without logarithmic
corrections is found for $d=2$ dimensions, but multiscaling arises for $d\ne 2$ \cite{Wald17}.} See also table~\ref{tab2}.

Logarithmic sub-ageing might also occur in the ageing of glassy materials. It is common to fit experimental data on relaxation phenomena in glasses through
sub-ageing, with a sub-ageing exponent $\mu\lesssim 1$, see \cite{Stru78,Vinc07} and refs. therein, 
which in practise may become very difficult to distinguish from a logarithmic subageing \cite{Bert00}.

Our discussion has been restricted to the special case $T=0$. The solution of the spherical constraint for $T>0$ is left as an open problem.
Since the conservation laws (\ref{2.8}) of the second model are maintained for $T\ne 0$, one might anticipate that a sufficiently small change
in $T$ should not lead to a drastic modification of the qualitative behaviour of the second model. In the third model, however, for any $T>0$
the conservation laws (\ref{2.8}) are broken and in consequence, the behaviour of the model should change notably. Another question left open
is the extension to any dimension $d$.

%%%%%%%%%%%%%%%%%%%%%%%%%%%  fin du texte principal %%%%%%%%%%%%%%%%%%%%%%%%%%%%%%%%%%%%%%%%%%%%%%%%%%%%%%%%%%%%%%%%%%%%%%%%%%%%%%%%%%%%%%%

\newpage

%
%%% Annexe
%
%%%%%%%%%%%%%%%%%%%%%%%%%%%%%%%%%%%%%%%%%%%%%%%%%%%%%%%%%%%%%%%%%%%%%%%%%%%%%%%%%%%%%%
\appsection{A}{The first Arcetri model revisited}            % jadis appele annexe E %
%%%%%%%%%%%%%%%%%%%%%%%%%%%%%%%%%%%%%%%%%%%%%%%%%%%%%%%%%%%%%%%%%%%%%%%%%%%%%%%%%%%%%%

We recall and extend the so-called `first' Arcetri model, as originally introduced in \cite{Henk15}, in order to clarify
the possible interpretations, either as a model of interface growth or else of interacting particles.
For brevity of notation, we restrict to $d=1$ dimensions.

A model of growing interfaces is naturally described in terms of the height $h_n(t)$
at the site $n$ of a periodic chain with $N$ sites.
The defining equation of motion of the {\em first Arcetri model for the heights} ({\sc Arcetri 1h}) is,
along with the spherical constraint on the slopes
\BEA
\partial_t h_n(t) &=& \nu \left( h_{n+1}(t) + h_{n-1}(t) - 2 h_n(t) \right) + \mathfrak{z}(t) h_n(t) +\eta_n(t) ~~~~~~
\label{E.1} \\
& & \frac{1}{4}\sum_{n=0}^{N-1} \left\langle\!\left\langle\, \left\langle
\left( h_{n+1}(t)- h_{n-1}(t)\right)^2 \right\rangle \right\rangle\!\right\rangle = N
\label{E.2}
\EEA
Herein, to each lattice site $n$ a centred gaussian random variable $\eta_n(t)$ is attached, with variance
$\left\langle \eta_n(t) \eta_m(t') \right\rangle = 2 \nu T \delta(t-t') \delta_{n,m}$. The
Lagrange multiplier $\mathfrak{z}(t)$ is determined from the `spherical constraint' (\ref{E.2})
and $\nu$ and $T$ are constants. A natural initial condition stipulates an initial gaussian distribution,
with spatially translation-invariant moments
\BEQ
\left\langle\!\left\langle h_n(0) \right\rangle\!\right\rangle = H_0 \;\; , \;\;
\left\langle\!\left\langle h_n(0) h_m(0) \right\rangle\!\right\rangle
- \left\langle\!\left\langle h_n(0) \right\rangle\!\right\rangle\left\langle\!\left\langle h_m(0) \right\rangle\!\right\rangle
= H_1\, \delta_{n,m}
\label{E.3}
\EEQ
This describes an initially flat interface of uncorrelated heights and of initial mean height $H_0$.

If one re-writes the equation of motion (\ref{E.1}) in terms of the slopes \\ 
$u_n(t) := \demi \left( h_{n+1}(t) - h_{n-1}(t) \right)$, one has
\BEA
\partial_t u_n(t) &=& \nu \left( u_{n+1}(t) + u_{n-1}(t) - 2 u_n(t) \right) + \mathfrak{z}(t) u_n(t)
+ \demi \left( \eta_{n+1}(t) - \eta_{n-1}(t) \right) ~~~~~~
\label{E.4} \\
&&\sum_{n=0}^{N-1} \left\langle\!\left\langle\, \left\langle u_n(t)^2 \right\rangle\, \right\rangle\!\right\rangle = N
\label{E.5}
\EEA
which together with the initial conditions (\ref{E.3}), with $H_0=0$, was the only model studied in \cite{Henk15}.
The formal continuum limit of (\ref{E.4}) is given by (\ref{Arc1}).
The {\em first Arcetri model for the particles} (or slopes) ({\sc Arcetri 1u}) has the defining equation of motion (\ref{E.4}),
the spherical constraint (\ref{E.5}) and an initial gaussian distribution, with the moments
\BEQ \label{E.6}
\left\langle\!\left\langle u_n(0) \right\rangle\!\right\rangle = 0 \;\; , \;\;
\left\langle\!\left\langle u_n(0) u_m(0) \right\rangle\!\right\rangle = U_1\, \delta_{n,m}
\EEQ
This initial condition with zero average slope corresponds to a system of initially uncorrelated
particles with average mean density $\vro=\demi$.

The solution of the equations of motion is standard, see \cite{Henk15}.
Let $g(t) := \exp\left( - 2 \int_0^{t}\!\D\tau\, \mathfrak{z}(\tau)\right)$, and define as well
\BEA
f(t) &:=& \frac{1}{2\pi} \int_{-\pi}^{\pi} \!\D p\: \sin^2 \!p \;\; e^{-4\nu(1-\cos p)t} \:=\: \frac{e^{-4\nu t}I_1(4\nu t)}{4\nu t}
\nonumber \\
F(t) &:=& \frac{1}{2\pi} \int_{-\pi}^{\pi} \!\D p\;\; e^{-4\nu(1-\cos p)t} \hspace{1.0truecm}\:=\: e^{-4\nu t}I_0(4\nu t)
\EEA
where the $I_n$ are modified Bessel functions \cite{Abra65}. For $d\geq 1$ dimensions, these functions become \cite{Henk15}
\BEQ
f(t) = d\frac{e^{-4\nu t}I_1(4\nu t)}{4\nu t} \left( e^{-4\nu t}I_0(4\nu t)\right)^{d-1} \;\; , \;\;
F(t) = \left( e^{-4\nu t}I_0(4\nu t)\right)^{d}
\EEQ
The spherical constraints (\ref{E.2},\ref{E.5})
reduce to the Volterra integral equations
\BEQ
\left\{
\begin{array}{ll}
g(t) = H_1 f(t) + 2\nu T \int_0^{t}\!\D\tau\: g(\tau)f(t-\tau)   & \mbox{\rm ~;~~ for {\sc Arcetri 1h} }  \\[0.15truecm]
g(t) = U_1 F(t) + 2\nu T \int_0^{t}\!\D\tau\: g(\tau)f(t-\tau)   & \mbox{\rm ~;~~ for {\sc Arcetri 1u} }
\end{array} \right.
\EEQ
which gives immediately for the Laplace transformation $\lap{g}(p)=\int_0^{\infty}\!\D t\: e^{-pt}\, g(t)$
\BEQ
\lap{g}(p) = \left\{
\begin{array}{ll} H_1 \lap{f}(p)/\left[1-2\nu T \lap{f}(p)\right]   & \mbox{\rm ~;~~ for {\sc Arcetri 1h} }  \\[0.15truecm]
                  U_1 \lap{F}(p)/\left[1-2\nu T \lap{f}(p)\right]   & \mbox{\rm ~;~~ for {\sc Arcetri 1u} }
\end{array} \right.
\EEQ
such that the small-$p$ behaviour of $\lap{g}(p)$ is related by a Tauberian theorem
to the long-time asymptotics of $g(t)$ for $t\to\infty$ \cite{Fell71}. This gives two distinct physical situations:

\noindent
{\bf (a)} If an interpretation in terms of \underline{interface growth} is sought, one may consider the {\sc Arcetri 1h} model,
characterised by initially uncorrelated height variables according to (\ref{E.3}). Then the time-dependent height is
\BEQ
\left\langle\!\left\langle \left\langle h_n(t) \right\rangle \right\rangle\!\right\rangle = H_0\, g(t)^{-1/2}
\EEQ
This average is indeed non-vanishing, since the equation of motion (\ref{E.1}) is not invariant under the
transformation $h_n(t) \mapsto h_n(t)+\alpha$. The two-time autocorrelator is given by
\BEA
C(t,s) &:=& \left\langle\!\left\langle\, \left\langle
\left(h_n(t) - \left\langle\!\left\langle \left\langle h_n(t) \right\rangle \right\rangle\!\right\rangle\right)
\left(h_n(s) - \left\langle\!\left\langle \left\langle h_n(s) \right\rangle \right\rangle\!\right\rangle\right)
\right\rangle\, \right\rangle\!\right\rangle
\nonumber \\
&=&\frac{H_1 F( (t+s)/2)}{\sqrt{ g(t) g(s)\,}\,}
+ 2\nu T \int_{0}^{\min(t,s)} \!\D\tau\: \frac{g(\tau)}{\sqrt{g(t)g(s)\,}\,}\, F\left(\frac{t+s}{2}-\tau\right)
\label{E.12}
\EEA
such that the interface width becomes
\BEQ
w^2(t) = C(t,t) = \frac{H_1 F(t)}{g(t)} + 2\nu T \int_0^t \!\D\tau\: \frac{g(\tau)}{g(t)}\, F(t-\tau)
\EEQ
Finally, the linear autoresponse of the height $h_n(t)$ with respect to a change $h_n(s) \mapsto h_n(s)+j_n(s)$
in the height is independent of the initial distribution and reads
\BEQ
R(t,s) := \left. \frac{\delta  \left\langle h_n(t) \right\rangle}{\delta j_n(s)}\right|_{j=0}
= \Theta(t-s) \sqrt{\frac{g(s)}{g(t)}\,}\, F\left(\frac{t-s}{2}\right)
\label{E.14}
\EEQ
where the Heaviside function $\Theta(t)$ expresses causality. Eqs.~(\ref{E.12},\ref{E.14}) are the analogues of (\ref{3.34},\ref{3.35}) in the
third model, for $n=0$.\footnote{In the {\sc kpz} universality class, the {\em {\sc kpz} ansatz} stipulates that 
$\left\langle\!\left\langle \left\langle h_n(t) \right\rangle \right\rangle\!\right\rangle - v_{\infty} t \sim t^{\beta}$ and 
$w(t)\sim t^{\beta}$ scale both with the same exponent $\beta$ \cite{Prae00}, and where
$v_{\infty}$ is the mean velocity of particle deposition 
(one may achieve $v_{\infty}=0$ by the choice of a co-moving frame of reference, implicit in (\ref{gl:KPZ-EW})). 
The {\sc kpz} ansatz is satisfied by the Arcetri 1H model at $T=T_c$, but does not hold for $T<T_c$. In the third Arcetri model at $T=0$, the
{\sc kpz} ansatz is broken through logarithmic sub-scaling exponents, see section~4.} 

\noindent
{\bf (b)} If a comparison with the \underline{$1D$ {\sc tasep}} is sought, one may consider the {\sc Arcetri 1u} model,
characterised by initially uncorrelated slopes and described by (\ref{E.6}).  The slope-slope auto-correlator $C(t,s)$, related to the
connected density-density correlator via (\ref{1.11}),
the linear auto-response $R(t,s)$ of the slope $u_n(t)$ with respect to a change
$u_n(s) \mapsto u_n(s) + k_n(s)$ in the slope, and the linear auto-response ${\cal R}(t,s)$ of the slope $u_n(t)$ with respect to a change
$j_n(s)$ in the height, respectively, are given by
\BEA
C(t,s) &:=& \left\langle\!\left\langle\, \left\langle u_n(t) u_n(s) \right\rangle \,\right\rangle\!\right\rangle\:=\:
\frac{U_1 F((t+s)/2)}{\sqrt{g(t)g(s)\,}\,}
+ 2\nu T \int_0^{\min (t,s)} \!\D\tau\: \frac{g(\tau)}{\sqrt{g(t)g(s)\,}\,}\, f\left(\frac{t+s}{2}-\tau\right)
\nonumber \\
R(t,s) &:=& \left.\frac{\delta \langle u_n(t)\rangle}{\delta k_n(s)}\right|_{k=0} \:=\:
\Theta(t-s) \sqrt{\frac{g(s)}{g(t)}\,}\, F\left(\frac{t-s}{2}\right)
\label{E.15} \\
{\cal R}(t,s) &:=& \left.\frac{\delta \langle u_n(t)\rangle}{\delta j_n(s)}\right|_{j=0} \:=\:
\Theta(t-s) \sqrt{\frac{g(s)}{g(t)}\,}\, f\left(\frac{t-s}{2}\right)
\nonumber
\EEA
Eq.~(\ref{E.15}) gives the analogues of (\ref{CorrA},\ref{RespQ},\ref{3.29}) in the second model.

Following the analysis in \cite{Henk15}, the critical exponents are readily found and are listed in tables~\ref{tab1} and~\ref{tab2}.

%%%%%%%%%%%%%%%%%%%%%%%%%%%%%%%%%%%%%%%%%%%%%%%%%%%%%%%%%%%%%%%%%%%%%%%%%%%%%%%%
\appsection{B}{Long-time correlator}                   % jadis appele annexe A %
%%%%%%%%%%%%%%%%%%%%%%%%%%%%%%%%%%%%%%%%%%%%%%%%%%%%%%%%%%%%%%%%%%%%%%%%%%%%%%%%
We derive the long-time behaviour of the correlations in the second and third model, at $T=0$.

In what follows, we shall often need the following asymptotic formula \cite{Abra65,Singh85}
\BEQ \label{A1}
I_n(x) \simeq \frac{1}{\sqrt{2\pi x\,}\,}\exp\left( x - \frac{4n^2-1}{8x} \right) \left( 1 + {\rm O}(x^{-2})\right)
\EEQ

First, we must find $Z(t)$ for large times from the constraint (\ref{3.15}) or (\ref{3.33}), respectively. We now prove eq.~(\ref{4.3}) in the main text.

For the second model, (\ref{3.15}) becomes (\ref{4.1}). Since the Bessel function $I_0(x)$ is monotonically
increasing with $x>0$, this implies that $Z(t)$ increases with $t>0$. Then, one can apply (\ref{A1}) and one has
\BEA
e^{4\nu t} &=& I_0\left( 4\nu t\sqrt{1+\frac{Z^2(t)}{4\nu^2 t^2}\,}\,\right) \simeq
\frac{\exp 4\nu t\sqrt{1+\frac{Z^2(t)}{4\nu^2 t^2}\,}\,}{\left[8\pi\nu t\sqrt{1+\frac{Z^2(t)}{4\nu^2 t^2}\,}\,\right]^{1/2}}
\nonumber \\
&\simeq & \exp\left[ 4\nu t \left(1+\frac{Z^2(t)}{8\nu^2 t^2}\right) -\demi\ln(8\pi\nu t) -\demi \ln\left( 1+\frac{Z^2(t)}{8\nu^2 t^2}\right)\right]
\nonumber
\EEA
Keeping the terms of leading non-vanishing order, gives a linear equation for $Z^2(t)$
\BD
\frac{Z^2(t)}{2\nu t} - \demi \ln(8\pi\nu t) = 0
\ED
equivalent to the first eq.~(\ref{4.3}) in the main text. One might obtain this result as well from the integral representation of $\mathscr{J}_0$, by the
saddle-point method.

For the third model, the constraint (\ref{3.33}) takes the form, using appendix~D and standard formul{\ae} for the modified Bessel function \cite{Abra65}
\BEA 
& & \left[ I_1\left(\sqrt{(4\nu t)^2 + (2 Z(t))^2}\:\right)(4\nu t)^2 \right. \nonumber \\
& & \left.+
2 Z(t)^2\sqrt{ (4\nu t)^2 + (2 Z(t))^2\,}\, \left( I_0\left(\sqrt{(4\nu t)^2 + (2 Z(t))^2}\:\right)+I_2\left(
\sqrt{(4\nu t)^2 + (2 Z(t))^2}\:\right)\right)\right] \nonumber \\
&=& e^{4\nu t} {\left[ (4\nu t)^2 + (2Z(t))^2\right]^{-3/2}}
\EEA
As before, we use (\ref{A1}) and expand to the lowest required order. We then find
\BEA
e^{4\nu t} {\left[ (4\nu t)^2 + (2Z(t))^2\right]^{-3/2}} 
= \frac{\exp 4\nu t\sqrt{1+\frac{Z^2(t)}{4\nu^2 t^2}\,}\,}{\left[8\pi\nu t\sqrt{1+\frac{Z^2(t)}{4\nu^2 t^2}\,}\,\right]^{1/2}} 
\left[ (4\nu t)^2 +(2 Z(t))^2 4\nu t \right] \nonumber
\EEA
which can be further simplified into
\BD
\frac{Z(t)^2}{2\nu t} - \demi\ln(2\pi) -\frac{7}{2} \ln\left( 4\nu t \right) +\ln\left( (4\nu t)^2 \left( 1 +\frac{(2Z(t))^2}{4\nu t} \right) \right)=0
\ED
The single positive solution of this equation is 
\BEQ \label{ZB3}
Z(t) = \sqrt{2t W\left( \left(32\pi e\right)^{1/2} t^{3/2}\right) -t\,}
\EEQ
where $W(x)$ denotes Lambert's $W$-function, defined as the solution of $W e^{W}=x$  \cite{Lambert1758,Corl96}.
Throughout, we shall require the following two expansions of Lambert's function
\BEQ
W(x) \simeq \left\{ \begin{array}{ll}
x - x^2 + {\rm O}(x^3) & \mbox{\rm ~~;~ for $x\to 0$} \\
\ln x - \ln(\ln x) + {\rm O}(\ln(\ln x)/\ln(x)) & \mbox{\rm ~~;~ for $x\to \infty$}
\end{array} \right.
\EEQ
Inserting into (\ref{ZB3}), we finally have the leading asymptotics for $Z(t)$, including the dominant logarithmic corrections
\BEQ
Z(t) \simeq \sqrt{3t\ln (\kappa_2 t) \left[ 1 - \frac{2}{3}\frac{\ln (\frac{3}{2} \ln \kappa_2 t)}{\ln \kappa_2 t} -\frac{1}{3}\frac{1}{\ln \kappa_2 t} \right]\,}
=: \sqrt{3t\ln (\kappa_2 t) \digamma(t)\,}
\EEQ
and we have derived eqs.~(\ref{4.3},\ref{4.4}) in the main text, and especially the values of $\kappa_1$ and $\kappa_2$ quoted therein.
For later use, below and in appendix~C, we also defined the function $\digamma(t)$. This function describes additional modifications of the scaling
behaviour of the third model with respect to the second model, where from the constraint (\ref{4.1}) we had seen that $\digamma(t)=1$. 

Since the abstract expression (\ref{CorrA}) holds true for both the slope correlator in the second model and the height correlator in the third model,
respectively, both can be analysed together. We begin with an analysis of the scaling behaviour of the autocorrelator, which reads
\BEA
C(t,s)= e^{-2\nu(t+s)} {I_0 \left(2\nu (t+s)\sqrt{1+ \left(\frac{Z(t)+Z(s)}{2\nu (t+s)}\right)^2\,}\:\right)}
\EEA
In what follows, we shall use the logarithmic subageing scaling variable
\BEQ \label{B7_scal} 
\tau = t-s = \frac{s}{\ln^{\vartheta}\kappa_2 s}(y-1) g(s)
\EEQ
which is to be considered in the double limit $t,s\to\infty$ such that $y>1$ is being kept fixed, and a positive constant $\vartheta>0$, 
in analogy with, but generalising \cite{Bert00}.
In certain cases, as we shall see especially in appendix~C when analysing the autoresponse function, the function $g(s)$ has to be conveniently chosen.
For what follows, an important simplification is obtained for the auxiliary function $\digamma(t)$. Inserting the
scaling ansatz and expanding, we find
\BEQ \label{B8}
\digamma(t) \simeq 1 - \frac{2\ln \frac{3}{2} -1}{6\ln\kappa_2 s} -\frac{1}{3}\frac{\ln\ln\kappa_2 s}{\ln \kappa_2 s} 
+{\rm O}(\ln^{-2}\kappa_2 s) \simeq \digamma(s)
\EEQ
to this order. 

Our first scaling analysis uses again (\ref{A1}), and (\ref{4.3}). We find by expansion, up to the first non-vanishing order, and using (\ref{B8})
\BEA
\ln C(t,s) &\simeq & \frac{\left( Z(t) + Z(s)\right)^2}{4\nu(t+s)} - \demi \ln(4\pi\nu (t+s))  +{\rm O}(s^{-1},t^{-1})
\nonumber \\
&\simeq & \digamma(s) \frac{\left( \sqrt{\kappa_1 \nu t\ln \kappa_2 t\,} 
+ \sqrt{\kappa_1\nu s\ln \kappa_2 s\,}\:\right)^2}{4\nu(t+s)} - \demi \ln(4\pi\nu (t+s))
\label{A.4}
\EEA
We use the scaling ansatz (\ref{B7_scal}). Inserting into the above, and expanding, we obtain
\BEA
\ln C(t,s) &\simeq& -\demi \ln(8\pi\nu) + \frac{\kappa_1}{2}\ln \kappa_2 +\demi(\kappa_1-1)\ln s
-\frac{\kappa_1}{32} (y-1)^2 \ln^{1-2\vartheta}(\kappa_2 s) \nonumber \\
& & - \delta_{\kappa_1,3}\left( \demi\ln\ln\kappa_2 s + \frac{1}{4}\left(2\ln\frac{3}{2}-1\right) \right) + {\rm o}(1) \nonumber
\EEA
where the contributions in the second line only arise for the third model. 
Multiscaling can only be avoided by choosing $\vartheta=\demi$ \cite{Bert00}.
This gives (\ref{Ascaling},\ref{Afonction}) in the main text and is
the first type of scaling behaviour of $C(t,s)$ to be considered.

However, different ways to obtain a scaling behaviour exist. These are found by considering the difference $\tau=t-s$ between the two times
and by making the change of variables
\BEQ \label{A.5}
\tau = t-s = \frac{s}{\ln^{\vartheta}(\kappa_2 s)} f(s,y)
\EEQ
where the unknown function $f=f(s,y)$ is assumed to be small compared to $\ln^{\vartheta} s$.
The function $f=f(s,y)$ must be found such that $\tau=\tau(y)$ is monotonically increasing with $y$ and the
autocorrelator $C=C(y)$ is monotonically decreasing.
Of course, the relation (\ref{A.5}) is to be understood in the scaling limit $t,s\to\infty$ with $y$ being kept fixed.
Using again (\ref{A.4}) and expanding as before, we have
\BEQ \label{A.6}
\ln C(t,s) \simeq \frac{\kappa_1-1-\delta_{\kappa_1,3}}{2}\ln(\ln \kappa_2 s) -\frac{\kappa_1}{32} \ln^{1-2\vartheta}(\kappa_2 s)\: f^2(s,y)
-\frac{\delta_{\kappa_1,3}}{4}\left(2\ln\frac{3}{2}-1\right)
\EEQ
In order to obtain a scaling behaviour, we make the following ansatz
\BEQ \label{A.7}
\ln C(t,s) \stackrel{!}{=} A'\ln(\ln\kappa_2 s) -\ln(y-1) +\demi \ln f^2(s,y) -\frac{B'}{32}\ln^{1-2\vartheta}(\kappa_2 s)\: f(s,y)
-\frac{\delta_{\kappa_1,3}}{4}\left(2\ln\frac{3}{2}-1\right)
\EEQ
where the constants $A',B'$ are to be determined. Consistency between (\ref{A.6}) and (\ref{A.7}) gives the condition
\BEQ
f^2 \exp\left[\frac{\kappa_1-B'}{16} \ln^{1-2\vartheta}(\kappa_2 s)\: f^2 \right] 
= (y-1)^2 \ln^{\kappa_1-2A'-1-\delta_{\kappa_1,3}}(\kappa_2 s)
\EEQ
which has the unique solution
\BEQ \label{A.9}
f^2 = \frac{16}{\kappa_1-B'} \ln^{1-2\vartheta}(\kappa_2 s)\: 
W\left(\frac{\kappa_1-B'}{16} (y-1)^2 \ln^{\kappa_1-2A'-2\vartheta-\delta_{\kappa_1,3}}(\kappa_2 s) \right)
\EEQ
using again the Lambert-$W$ function. Herein, self-consistency requires
that $\kappa_1-B'>0$ and $\kappa_1-2A'-2\vartheta-\delta_{\kappa_1,3}<0$. 
Then, we find the following asymptotic form, for $s\to\infty$
\BEQ \label{A.10}
\tau = t-s = \frac{s}{\ln^{(1+\delta_{\kappa_1,3}-\kappa_1)/2+A'+\vartheta}(\kappa_2 s)}\, (y-1)
\EEQ
and we see that at least for the autocorrelator, we can simply set $g(s)=1$. 
This explains the chosen ansatz: we have chosen variables such that in the special case 
where $(1+\delta_{\kappa_1,3}-\kappa_1)/2+A'+\vartheta=0$, we recover the scaling
form $\tau=t-s=s(y-1)$ of simple ageing. This discussion will be completed by a comparison 
with the corresponding results from the autoresponse $R(t,s)$, see appendix~C.

In order to derive (\ref{CorrAinf}), we reuse eq.~(\ref{A1}) in (\ref{CorrA}) 
and also recall that for large times, $Z(t)\sim t^{1/2}$, up to logarithmic
factors. All factors which do not contain $n$ will be absorbed into the autocorrelator $C(t,s)$.
Therefore, the leading $n$-dependent term coming from the Bessel function $I_n$ in 
(\ref{CorrA}) is simply $\exp\left[-n^2/(4\nu(t+s))\right]$, whereas
those terms which contain $Z(t)$ or $Z(s)$  will give rise to finite-time 
corrections to the leading scaling contribution. Hence (\ref{CorrAinf}) describes
the leading scaling behaviour of the time-space correlator $C_n(t,s)$.

In order to derive the spatial modulation of the time-space correlator (\ref{ACorrel2}), 
we start from (\ref{CorrAinf}). For large times $t,s\to\infty$,
the modulating factor can be rewritten as follows
\BD
\cos\left(n \frac{Z(t)+Z(s)}{2\nu(t+s)} \right) = \cos\left( \frac{n}{\sqrt{2\nu(t+s)}\,}\sqrt{\frac{(Z(t)+Z(s))^2}{2\nu(t+s)}\,}\:\right)
\ED
Herein, since $Z(t)\sim t^{1/2}$, we used that the argument of the $\arctan$ is small so that it is enough to keep the lowest order. Now, straightforward
expansion of the square root produces the stated form (which is symmetric in $t$ and $s$) and which can be done, up to finite-time corrections.

Finally, the single-time correlator $C_n(t,t)=\lim_{s\to t} C_n(t,s)$ is read off immediately from (\ref{ACorrel2}) to produce (\ref{ACorrel1}).

%%%%%%%%%%%%%%%%%%%%%%%%%%%%%%%%%%%%%%%%%%%%%%%%%%%%%%%%%%%%%%%%%%%%%%%%%%%%%%%%
\appsection{C}{Long-time response}                  % jadis appele annexe B    %
%%%%%%%%%%%%%%%%%%%%%%%%%%%%%%%%%%%%%%%%%%%%%%%%%%%%%%%%%%%%%%%%%%%%%%%%%%%%%%%%

The analysis of the autoresponse starts from
\BEA
R(t,s)= e^{-2\nu(t-s)} {I_0 \left(2\nu (t-s)\sqrt{1+ \left(\frac{Z(t)-Z(s)}{2\nu (t-s)}\right)^2}\right)}
\EEA
Expanding the Bessel function via (\ref{A1}) and using (\ref{4.3},\ref{4.4},\ref{B7_scal}), we obtain
\BEA
\ln R(t,s) &\simeq & \frac{\left( Z(t) - Z(s)\right)^2}{4\nu(t-s)} - \demi\ln(4\pi\nu(t-s)) +{\rm O}(s^{-1},t^{-1})
\nonumber \\
&\simeq & \digamma(s)\frac{\left( \sqrt{\kappa_1\nu t\ln \kappa_2 t} - \sqrt{\kappa_1\nu s\ln \kappa_2 s}\right)^2}{4(t-s)} -
\demi \ln(4\pi\nu(t-s))
\label{B.2}
\EEA
In analogy with our analysis for the autocorrelator in appendix~B, we try to find  a scaling variable $y$ such that the time difference $\tau=\tau(y)$
increases monotonically with $y$ and that the response $R=R(y)$ decreases monotonically with $y$. The time difference is written as
\BEQ
\tau = t-s = \frac{s}{\ln(\kappa_2 s)} f(s,y)
\EEQ
where the unknown function $f=f(s,y)$ plays the r\^ole of the scaling variable. We assume that $f\ln s\ll 1$.
Then we can expand $R(t,s)$. After several cancellations, we finally arrive for the second model and third model, respectively, at
\begin{subequations} \label{B.3}
\begin{align} 
\ln R(t,s) &\simeq \frac{\kappa_1}{16}f(s,y) -\demi\ln f(s,y) -\demi\ln (4\pi\nu s) +\demi \ln(\ln\kappa_2 s) + {\rm o}(1) \label{B.3a} \\
\ln R(t,s) &\simeq \frac{f(s,y)}{16}\left(\kappa_1  - \frac{\ln\ln\kappa_2 s}{\ln \kappa_2 s}\right)
-\demi\ln f(s,y) -\demi\ln (4\pi\nu s) +\demi \ln(\ln\kappa_2 s) + {\rm o}(1) \nonumber \\
&= \frac{\kappa_1\bar{f}(s,y)}{16} -\demi\ln \bar{f}(s,y) -\demi\ln\left( 1 -\frac{\ln\ln\kappa_2 s}{\kappa_1\ln \kappa_2 s}\right)
-\demi\ln (4\pi\nu s) +\demi \ln(\ln\kappa_2 s) + {\rm o}(1) \nonumber \\
&\simeq \frac{\kappa_1\bar{f}(s,y)}{16} -\demi\ln \bar{f}(s,y)  
-\demi\ln (4\pi\nu s) +\demi \ln(\ln\kappa_2 s) + {\rm o}(1) \label{B.3b}
\end{align}
\end{subequations}
where for the third model we redefined the scaling variable 
$\bar{f}(s,y) := f(s,y) \left(1  - \frac{\ln\ln\kappa_2 s}{\kappa_1\ln \kappa_2 s}\right)\stackrel{s\to\infty}{\simeq} f(s,y)$.
The scaling of the autoresponse of the second model in (\ref{B.3a}) and of the third model in (\ref{B.3b}) can be discussed
simultaneouly, by using the scaling variables $f$ or $\bar{f}$, respectively. 

Now, we can {\em define} a scaling variable $y>1$, for $s\to\infty$, through the ansatz (using $f$ or $\bar{f}$, respectively) 
\BEQ \label{B.4}
\ln R(t,s) \stackrel{!}{=} -\demi\ln(4\pi\nu s) -\demi\ln (y-1) + \frac{B}{16} f(s,y) + A \ln(\ln \kappa_2 s)
\EEQ
where $A,B$ are constants. Consistency of (\ref{B.3},\ref{B.4}) leads to
\BEQ
f(s,y) = \frac{8}{B-\kappa_1}\, W\left( \frac{B-\kappa_1}{8}\, (y-1) \ln^{1-2A}(\kappa_2 s) \right)
\EEQ
where $2A>1$, $B-\kappa_1>0$ and using again Lambert's function $W(x)$. For the third model, one simply reads $\bar{f}$ instead of $f$. 
The response function becomes
\BD
\ln R(t,s) \simeq -\demi \ln(4\pi\nu s) + A\ln(\ln \kappa_2 s) - \demi\ln (y-1) +\frac{B}{16}
\underbrace{~~\ln^{1-2A}(\kappa_2 s)~~}_{\mbox{\rm $\to 0$ for $s\to\infty$}} (y-1)
\ED
and we have the final scaling form, with $A>\demi$
\BEQ
R(t,s) = (4\pi\nu s)^{-1/2} \ln^A(\kappa_2 s) \; (y-1)^{-1/2}
\EEQ
with the scaling variable
\BEQ \label{B.8}
t-s = \frac{s}{\ln\kappa_2 s} \frac{8}{B-\kappa_1} W\left( \frac{B-\kappa_1}{8}\, (y-1) \ln^{1-2A}(\kappa_2 s)\right)
\EEQ
such that for $s\to\infty$, we recover $t-s\simeq s \ln^{-2A}(\kappa_2 s)\, (y-1)$ with $2A>1$.

%%++++++++++++++++++++++++++++++++++++++++++++++++++++++++++++++++++++++++++++++++
%%++++++++++++++++++++++++++++++++++++++++++++++++++++++++++++++++++++++++++++++++

To finish the argument, we now compare with the scaling of the autocorrelator, discussed in appendix~B. Because of the condition $2A>1$, the
scaling (\ref{B.8}) cannot be compatible with (\ref{Ascaling}). {\it A contrario}, compatibility with (\ref{Cscaling}) can be achieved via the condition
\BEQ
2A = \frac{1+\delta_{\kappa_1,3}-\kappa_1}{2} + A' + \vartheta
\EEQ
It follows that the condition $2A>1$ implies the bound $\kappa_1-2A'-2\vartheta-\delta_{\kappa_1,3}<0$ obtained in appendix~B.
Since we shall only encounter the combination $A'+\vartheta$, we may as well fix one of those two constants. For example, we might insist
that the function $f$ in (\ref{A.9}) should not become singular for $s\to\infty$. This fixes $\kappa_1-2A'-2\vartheta-\delta_{\kappa_1,3}=0$, hence
\BEQ
\vartheta= 2A \;\; , \;\; A'=\demi(\kappa_1-1-\delta_{\kappa_1,3})
\EEQ
This produces the final forms (\ref{Cscaling}) and (\ref{Qscaling}) in the text, if we choose $B=\kappa_1+8$ and $B'=\kappa_1-16$.
In figure~\ref{fig2} we compare the functions $\tau=\tau(s,y)$,
for $s$ finite and fixed, for responses and correlators. This illustrates that unless $s$ becomes enormously large, there are strong non-linearities in the
scaling variable to be taken into account.

Finally, the asymptotic form in (\ref{QCorrel2}) is derived in completely analogy with the treatment of the correlator in appendix~B.
The absence of a spatial modulation in (\ref{QCorrel2}) follows from 
\BEQ
\frac{\left( Z(t) - Z(s)\right)^2}{2\nu(t-s)} = {\rm O}(\ln^{-1} s)
\EEQ

%\newpage
%%%%%%%%%%%%%%%%%%%%%%%%%%%%%%%%%%%%%%%%%%%%%%%%%%%%%%%%%%%%%%%%%%%%%%%%%%%%%%%%
\appsection{D}{On some special functions}              % jadis appele annexe F %
%%%%%%%%%%%%%%%%%%%%%%%%%%%%%%%%%%%%%%%%%%%%%%%%%%%%%%%%%%%%%%%%%%%%%%%%%%%%%%%%

We compute the functions $\mathscr{J}_a$, with $a\in\mathbb{N}$, defined by
\BEA
\mathscr{J}_{2a}(A,Z) &:=& \frac{1}{\pi} \int_{0}^{\pi} \!\D k\: e^{A\cos k} \cosh( Z \sin k) \left( \sin k\right)^{2a} =
\frac{\partial^{2a}}{\partial Z^{2a}} \mathscr{J}_0(A,Z)
\\
\mathscr{J}_{2a+1}(A,Z) &:=& \frac{1}{\pi} \int_{0}^{\pi} \!\D k\: e^{A\cos k} \sinh( Z \sin k) \left( \sin k\right)^{2a+1} =
\frac{\partial^{2a+1}}{\partial Z^{2a+1}} \mathscr{J}_0(A,Z)
\EEA
and in principle, it is enough to find $\mathscr{J}_0$ explicitly. This can be done as follows
\BEA
\mathscr{J}_0(A,Z) &=& \frac{1}{\pi}\int_{0}^{\pi} \!\D k\: e^{A \cos k} \sum_{r=0}^{\infty} \frac{Z^{2r}}{(2r)!} \sin^{2r} k
\nonumber \\
&=& \sum_{r=0}^{\infty} \frac{1}{\pi} \frac{Z^{2r}}{\Gamma(2r+1)} \sqrt{\pi\,}\;
\Gamma\left(r+\demi\right) \left(\frac{A}{2}\right)^{-r} I_r(A)
\nonumber \\
&=& \sum_{r=0}^{\infty} \left(\frac{Z^2}{2A}\right)^r \frac{1}{r!} I_r(A)
\:=\: \sum_{r=0}^{\infty} \left(\frac{A}{2}\right)^r \frac{I_r(A)}{r!}\:
\left[ \left(\frac{Z}{A}\right)^2 \right]^r
\nonumber \\
&=& I_0\left( A \sqrt{ 1 + Z^2/A^2\,}\,\right) \:=\: I_0\left( \sqrt{A^2 + Z^2\,}\,\right)
\EEA
Herein, after expanding the $\cosh$ in the first line, we used an integral representation
\cite[eq. (9.6.18)]{Abra65} of the modified Bessel function
$I_r(A)$ in  the second line, simplified in the third line with the help of the duplication formula
\cite[eq. (6.1.18)]{Abra65} of the Gamma-function
and in the forth line applied the multiplication formula \cite[eq. (9.6.51)]{Abra65} for the $I_r$ (see also (\ref{G7}) below).

We notice the unexpected rotation-symmetry, in the $(A,Z)$-plane, of the integral identity
\BEQ
\frac{1}{\pi}\int_{0}^{\pi} \!\D k\: e^{A \cos k} \cosh( Z \sin k) = I_0\left( \sqrt{A^2 + Z^2}\,\right)
\EEQ
just derived.

Similarly, one can express $\mathscr{J}_1$, with the help of \cite[eqs. (9.6.18, 6.1.18, 9.6.51))]{Abra65} as :
\BEA
\mathscr{J}_1(A,Z) &=& \frac{1}{\pi}\int_{0}^{\pi} \!\D k\: e^{A \cos k} \sum_{r=0}^{\infty} \frac{Z^{2r}}{(2r)!} \sin^{2r+1} k
\nonumber \\
&=&  \sum_{r=0}^{\infty} \left(\frac{A}{2}\right)^r  \left(\frac{Z}{A}\right)^{2r+1}  \frac{I_{r+1}(A)}{r!}
\nonumber \\
&=& \frac{Z I_1\left( \sqrt{A^2+Z^2}\,\right)}{\sqrt{A^2+Z^2}\,}
\EEA
and one can also verify that $\mathscr{J}_1(A,Z)=\partial_Z \mathscr{J}_0(A,Z)$.
Similarly, $\mathscr{J}_2(A,Z)=\partial_Z \mathscr{J}_1(A,Z)$.

\newpage
%%%%%%%%%%%%%%%%%%%%%%%%%%%%%%%%%%%%%%%%%%%%%%%%%%%%%%%%%%%%%%%%%%%%%%%%%%%%%%%%
\appsection{E}{Proof of an identity}                  % jadis appele annexe G  %
%%%%%%%%%%%%%%%%%%%%%%%%%%%%%%%%%%%%%%%%%%%%%%%%%%%%%%%%%%%%%%%%%%%%%%%%%%%%%%%%

We derive the following identity, for all $n\in\mathbb{N}$ and $A,Z\in\mathbb{C}$ %{\small (provided the integral converges)}
\BEQ \label{G1}
\mathscr{C}_n(A,Z) := \frac{1}{\pi}\int_{0}^{\pi} \!\D k\: e^{A\cos k}
\cosh(Z \sin k) \cos(n k) = I_n\left( \sqrt{A^2 + Z^2}\,\right)
\cos \left( n \arctan \left(\frac{Z}{A}\right)\right)
\EEQ
where $I_n$ is a modified Bessel function. The special case $n=0$ is the function $\mathscr{J}_0(A,Z)$ derived in appendix~D.
For the proof, we shall require the following result:

\noindent {\bf Lemma:} {\it For any integers $n,\nu\in\mathbb{N}$ and $z,\mathfrak{z}\in\mathbb{C}$, one has}
\BEA
\frac{1}{\pi} \int_0^{\pi} \!\D\theta\: \exp(z\cos\theta) \cos(n\theta) \sin^{2\nu} \theta &=&
\frac{(-1)^{\nu}}{2^{2\nu}} \sum_{k=0}^{2\nu} (-1)^k \left(\vekz{2\nu}{k}\right) I_{n+2\nu-2k}(z) \label{G2} \\
\frac{1}{\pi} \int_0^{\pi} \!\D\theta\: \exp(\II\mathfrak{z}\cos\theta) \cos(n\theta) \sin^{2\nu} \theta &=&
\frac{\II^n}{2^{2\nu}} \sum_{k=0}^{2\nu} \left(\vekz{2\nu}{k}\right) J_{n+2\nu-2k}(\mathfrak{z})  \label{G3}
\EEA
{\it where $J_n$ is a Bessel function and $I_n$ is a modified Bessel function \cite{Abra65}.}

\noindent {\footnotesize {\bf Corollary:}  {\it Separating the real and imaginary parts in (\ref{G3}),
for $n$ even and odd, respectively, gives for $\mathfrak{z}\in\mathbb{R}$} ($m, \nu\in\mathbb{N}$)
\BEA
\frac{1}{\pi} \int_0^{\pi} \!\D\theta\: \cos(\mathfrak{z}\cos\theta) \cos(2m\theta) \sin^{2\nu} \theta &=&
\frac{(-1)^m}{2^{2\nu}} \sum_{k=0}^{2\nu} \left(\vekz{2\nu}{k}\right) J_{2m+2\nu-2k}(\mathfrak{z})  \nonumber \\
\frac{1}{\pi} \int_0^{\pi} \!\D\theta\: \sin(\mathfrak{z}\cos\theta) \cos((2m+1)\theta) \sin^{2\nu} \theta &=&
\frac{(-1)^m}{2^{2\nu}} \sum_{k=0}^{2\nu} \left(\vekz{2\nu}{k}\right) J_{2m+1+2\nu-2k}(\mathfrak{z})  \nonumber \\
\int_0^{\pi} \!\D\theta\: \sin(\mathfrak{z}\cos\theta) \cos(2m\theta) \sin^{2\nu} \theta &=&
\int_0^{\pi} \!\D\theta\: \cos(\mathfrak{z}\cos\theta) \cos((2m+1)\theta) \sin^{2\nu} \theta \:=\: 0 \nonumber
\EEA
}

\noindent
The strategy of proof will be as follows: (\ref{G1}) follows from (\ref{G2}),
which in turn is an immediate consequence of (\ref{G3}).

\noindent \underline{\bf Step 1:} To prove (\ref{G3}), recall Euler's formula,
$e^{\II\mathfrak{z}\cos\theta}=\cos(\mathfrak{z}\cos\theta) + \II \sin(\mathfrak{z}\cos\theta)$, with $\mathfrak{z}\in\mathbb{C}$.
Multiplying with $\cos n\theta$ and integrating gives, see \cite[eq. (9.1.21)]{Abra65}, with $n\in\mathbb{N}$
\BEQ \label{G4}
\II^n J_n(\mathfrak{z}) = \frac{1}{\pi} \int_0^{\pi} \!\D\theta\: \cos(\mathfrak{z}\cos\theta)\cos(n\theta)
+ \frac{\II}{\pi}\int_0^{\pi} \!\D\theta\: \sin(\mathfrak{z}\cos\theta)\cos(n\theta)
\EEQ
Next, denote the integral on the left-hand-side of (\ref{G3}) as
$C_{n,\nu}(\mathfrak{z}) := \frac{1}{\pi} \int_0^{\pi}\!\D\theta\: e^{\II\mathfrak{z}\cos\theta} \cos(n\theta) \sin^{2\nu}\theta$.
It is easily verified that one has the differential recurrence relation
\BEQ \label{G5}
C_{n,\nu+1}(\mathfrak{z}) = \partial_{\mathfrak{z}}^2 C_{n,\nu}(\mathfrak{z}) + C_{n,\nu}(\mathfrak{z})
\EEQ
For a fixed $n\in\mathbb{N}$, the identity (\ref{G3}) is the assertion that for all $\nu\in\mathbb{N}$ one has the identity:
\BEQ \label{G6}
C_{n,\nu}(\mathfrak{z}) = \frac{\II^n}{2^{2\nu}} \sum_{k=0}^{2\nu} \left(\vekz{2\nu}{k}\right) J_{n+2\nu-2k}(\mathfrak{z})
\EEQ
which we now prove by induction over $\nu$. For $\nu=0$, the assertion (\ref{G6}) is just the relation (\ref{G4}).
For the induction step $\nu\mapsto\nu+1$, we use\footnote{In this calculation, we write $J_n$ instead of fully $J_n(\mathfrak{z})$.}
(\ref{G5}) and apply first the Bessel function identities \cite[eq. (9.1.27)]{Abra65}
and then several times standard identities of the binomial coefficients
\BEA
\lefteqn{C_{n,\nu+1}(\mathfrak{z}) = \left( 1 + \frac{\partial^2}{\partial\mathfrak{z}^2} \right) C_{n,\nu}(\mathfrak{z}) } \nonumber \\
&=& \frac{\II^n}{2^{2\nu}}\sum_{k=0}^{2\nu} \left(\vekz{2\nu}{k}\right)
\left[ J_{n+2\nu-2k} +\frac{1}{4}\left[ J_{n-2+2\nu-2k}-2J_{n+2\nu-2k} +J_{n+2+2\nu-2k} \right]\right]
\nonumber \\
&=& \frac{\II^n}{2^{2(\nu+1)}}\sum_{k=0}^{2\nu} \left(\vekz{2\nu}{k}\right)
\left[ \stackrel{\stackrel{~}{}}{J}_{n+2\nu-2k} +J_{n-2+2\nu-2k}  +J_{n+2+2\nu-2k} +J_{n+2\nu-2k}\right]
\nonumber \\
&=& \frac{\II^n}{2^{2(\nu+1)}} \left\{
\sum_{k=0}^{2\nu} \left(\vekz{2\nu}{k}\right) J_{n+2\nu-2k}
+ \sum_{k=1}^{2\nu+1} \left(\vekz{2\nu}{k-1}\right) J_{n+2\nu-2k} \right. \nonumber \\
& & \left.
+ \sum_{k=0}^{2\nu} \left(\vekz{2\nu}{k}\right) J_{n+2(\nu+1)-2k}
+ \sum_{k=1}^{2\nu+1} \left(\vekz{2\nu}{k-1}\right) J_{n+2(\nu+1)-2k} \right\}
\nonumber \\
&=& \frac{\II^n}{2^{2(\nu+1)}} \left\{
J_{n+2\nu} + \sum_{k=1}^{2\nu} \left(\vekz{2\nu+1}{k}\right) J_{n+2\nu-2k} + J_{n+2\nu-2(2\nu+1)} \right. \nonumber \\
& & \left.
+ J_{n+2(\nu+1)} + \sum_{k=1}^{2\nu} \left(\vekz{2\nu+1}{k}\right) J_{n+2(\nu+1)-2k} + J_{n+2(\nu+1)-2(2\nu+1)} \right\}
\nonumber \\
&=& \frac{\II^n}{2^{2(\nu+1)}} \left\{
\stackrel{\stackrel{\stackrel{\stackrel{~}{}}{}}{}}{J}_{n+2\nu} + J_{n-2(\nu+1)} + J_{n+2(\nu+1)} + J_{n-2\nu} \right. \nonumber \\
& & \left. + \sum_{k=2}^{2\nu+1} \left(\vekz{2\nu+1}{k-1}\right) J_{n+2(\nu+1)-2k}
+ \sum_{k=1}^{2\nu} \left(\vekz{2\nu+1}{k}\right) J_{n+2(\nu+1)-2k} \right\}
\nonumber \\
&=& \frac{\II^n}{2^{2(\nu+1)}} \left\{
\stackrel{\stackrel{\stackrel{~}{}}{}}{J}_{n+2\nu} + J_{n+2(\nu+1)} + J_{n-2(\nu+1)} + J_{n-2\nu} +(2\nu+1) J_{n+2\nu} +(2\nu+1)J_{n-2\nu} \right.
\nonumber \\
& & \left. + \sum_{k=2}^{2\nu} \left(\vekz{2\nu+2}{k}\right) J_{n+2(\nu+1)-2k} \right\}
\nonumber \\
&=& \frac{\II^n}{2^{2(\nu+1)}}  \sum_{k=0}^{2(\nu+1)} \left(\vekz{2(\nu+1)}{k}\right) J_{n+2(\nu+1)-2k}(\mathfrak{z})  \nonumber
\EEA
which proves the assertion (\ref{G6}) for all $\nu\in\mathbb{N}$ (in the last line, we restored the argument $\mathfrak{z}$).

\noindent \underline{\bf Step 2:} starting from (\ref{G3}),
it is enough to set $\mathfrak{z}=\II z$ and to recall that $J_n(\II z) = \II^n I_n(z)$.
Eq.~(\ref{G3}) then gives
\BD
\frac{1}{\pi}\int_0^{\pi}\!\D\theta\: e^{-z\cos\theta} \cos(n\theta)\sin^{2\nu}\theta
= \frac{(-1)^{n+\nu}}{2^{2\nu}} \sum_{k=0}^{2\nu} \left(\vekz{2\nu}{k}\right) (-1)^k I_{n+2\nu-2k}(z).
\ED
Going over to $z\mapsto -z$, along with $I_n(-z)=(-1)^n I_n(z)$, produces (\ref{G2}). This proves the lemma.

\noindent \underline{\bf Step 3:} For proving (\ref{G1}), we require
two more preparations. First, recall the identity \cite[eq. (9.6.51)]{Abra65},
for $n\in\mathbb{N}$, $x\in\mathbb{C}$ and $\lambda\ne 0$
\BEQ \label{G7}
\sum_{\ell=0}^{\infty} \frac{(\lambda^2-1)^{\ell} (x/2)^{\ell}}{\ell !}\, I_{n\pm\ell}(x) = \lambda^{\mp n} I_{n}(\lambda x)
\EEQ
Second, let $x=\tan \vph$. Then
\BEQ \label{G8}
\left( \frac{1-\II x}{1+\II x}\right)^{n/2} + \left( \frac{1+\II x}{1-\II x}\right)^{n/2}
= \exp\left( -2\II\vph \frac{n}{2}\right) + \exp\left( +2\II\vph \frac{n}{2}\right)
= 2 \cos n\vph
\EEQ
Now, denote the left-hand-side of (\ref{G1}) by $\mathscr{C}_n=\mathscr{C}_n(A,Z)$.
Expanding the $\cosh$ in the integral representation of $\mathscr{C}_n$, we have
\BEA
\mathscr{C}_n &=& \frac{1}{\pi} \sum_{m=0}^{\infty} \frac{Z^{2m}}{(2m)!} \int_0^{\pi} \!\D k\: e^{A\cos k} \sin^{2m} k \cos (kn)
\nonumber \\
&=& \sum_{m=0}^{\infty} \sum_{\ell=0}^{2m} \left( \frac{Z}{2}\right)^{2m} \frac{(-1)^{m+\ell}}{(2m-\ell)! \ell !}\, I_{n+2m-2\ell}(A)
\nonumber \\
&=& \sum_{m=0}^{\infty} \sum_{\ell=0}^{m} \demi \left[ \stackrel{~}{1} + (-1)^{\stackrel{~}{m}}\right]
\left( \frac{Z}{2}\right)^{m} \frac{\II^m (-1)^{\ell}}{(m-\ell)! \ell !}\, I_{n+m-2\ell}(A)
\nonumber
\EEA
where in the second line, we used (\ref{G2}). In the last line, we replaced the even integer
$2m$ by the integer $m\in\mathbb{N}$, where
the extra factor guarantees that only the even values of $m$ give a non-vanishing contribution.
Now, we can exchange the order of summation and perform afterwards a shift in the summation variable $m$, to obtain
\BEA
\mathscr{C}_n &=& \sum_{\ell=0}^{\infty} \sum_{m=\ell}^{\infty} \demi
\left[ \stackrel{~}{1} + (-1)^{\stackrel{~}{m}}\right] \left( \frac{Z}{2}\right)^{m}
\frac{\II^m (-1)^{\ell}}{(m-\ell)! \ell !}\, I_{n+m-2\ell}(A)
\nonumber \\
&=& \demi \sum_{\ell=0}^{\infty} \sum_{m=0}^{\infty} \left[ (-1)^{\stackrel{~}{\ell}} + (-1)^{\stackrel{~}{m}}\right]
\left(\frac{\II Z}{2}\right)^{m+\ell} \frac{1}{m!\, \ell !}\, I_{n-\ell +m}(A)
\nonumber \\
&=& \demi \sum_{\ell=0}^{\infty} \frac{1}{\ell !} \left( \frac{\II Z}{2}\right)^{\ell} \sum_{m=0}^{\infty}
\left[ (-1)^{\stackrel{~}{\ell}} + (-1)^{\stackrel{~}{m}}\right] \frac{1}{m!}
\left(\frac{\II Z}{A}\right)^{m} \left(\frac{A}{2}\right)^m \, I_{(n-\ell)+m}(A)
\nonumber \\
&=& \demi \sum_{\ell=0}^{\infty} \frac{1}{\ell !} \left( \frac{\II Z}{2}\right)^{\ell}
\left[ (-1)^{\ell} \left( 1 + \II\frac{Z}{A}\right)^{-(n-\ell)/2}\, I_{n-\ell}\left( A \sqrt{1+\II\frac{Z}{A}\,}\,\right) \right.
\nonumber \\
& & \left.
~~~~~+  \left( 1 - \II\frac{Z}{A}\right)^{-(n-\ell)/2}\, I_{n-\ell}\left( A \sqrt{1-\II\frac{Z}{A}\,}\,\right) \right]
\nonumber \\
&=& \demi \left( 1 + \II\frac{Z}{A}\right)^{-n/2} \sum_{\ell=0}^{\infty} \frac{1}{\ell !}
\left( -\II\frac{Z}{A}\right)^{\ell} \left( \demi A \sqrt{1+\II\frac{Z}{A}\,}\,\right)^{\ell}\, I_{n-\ell}\left( A \sqrt{1+\II\frac{Z}{A}\,}\,\right)
\nonumber \\
& & + \demi \left( 1 - \II\frac{Z}{A}\right)^{-n/2} \sum_{\ell=0}^{\infty} \frac{1}{\ell !}
\left( \II\frac{Z}{A}\right)^{\ell} \left( \demi A \sqrt{1-\II\frac{Z}{A}\,}\,\right)^{\ell}\, I_{n-\ell}\left( A \sqrt{1-\II\frac{Z}{A}\,}\,\right)
\nonumber \\
&=& \demi \left[ \left( \frac{1-\II Z/A}{1+\II Z/A}\right)^{n/2} + \left( \frac{1+\II Z/A}{1-\II Z/A}\right)^{n/2} \right] \,
I_n\left( A \sqrt{1+\frac{Z^2}{A^2}\,}\,\right)
\nonumber \\
&=& \cos\left(n \arctan\frac{Z}{A} \right)\, I_n\left(\sqrt{A^2 + Z^2\,}\,\right)
\nonumber
\EEA
as asserted. In the calculation, we applied in the third and fifth lines the identity (\ref{G7}),
to carry out, first the sum over $m$, and then over $\ell$, and finally used (\ref{G8}) in the seventh line.

%%%%%%%%%%%%%%%%%%%%%%%%%%%%%%%%%%%%%%%%%%%%%%%%%%%%%%%%%%%%%%%%%%%%%%%%%%%%%%%%%%%%%%%
\appsection{F}{Discrete cosine- and sine-transformations}     % jadis appele annexe Z %
%%%%%%%%%%%%%%%%%%%%%%%%%%%%%%%%%%%%%%%%%%%%%%%%%%%%%%%%%%%%%%%%%%%%%%%%%%%%%%%%%%%%%%%

For the convenience of the reader, we recall some basic properties of discrete cosine and sine transformations.
On a periodic chain with $N$ sites, the cosine-transformation $\cal C$ of an even function
$a_n(t)=a_{-n}(t)$ is defined as, with $k=0,1,\ldots,N-1$
\BEQ
\wht{a}(t,k) = {\cal C}(a_n(t))(k) := \sum_{n=0}^{N-1} \cos\left(\frac{2\pi}{N} kn\right) a_n(t)
\;\; , \;\;
a_n(t) = \frac{1}{N}  \sum_{k=0}^{N-1} \cos\left(\frac{2\pi}{N} kn\right) \wht{a}(t,k)
\EEQ
and is itself even, viz. $\wht{a}(t,k)=\wht{a}(t,-k)$.
The sine-transformation $\cal S$ of an odd function $b_n(t)=-b_{-n}(t)$ is defined as
\BEQ
\wht{b}(t,k) = {\cal S}(b_n(t))(k) := \sum_{n=0}^{N-1} \sin\left(\frac{2\pi}{N} kn\right) b_n(t)
\;\; , \;\;
b_n(t) = \frac{1}{N}  \sum_{k=0}^{N-1} \cos\left(\frac{2\pi}{N} kn\right) \wht{b}(t,k)
\EEQ
and is itself odd, viz. $\wht{b}(t,k)=-\wht{b}(t,-k)$. Clearly, $\cal C$ and $\cal S$ are linear operators.
Furthermore, ${\cal C}(b_n(t))={\cal S}(a_n(t))=0$ and
${\cal C}^2(a_n(t)) = N a_n(t)$ and ${\cal S}^2(b_n(t)) = N b_n(t)$.

In the main text, we shall need frequently the following cosine-transformations of the even functions
\BEA
{\cal C}\left(a_{n+1}(t)+a_{n-1}(t)-2a_n(t)\right)(k) &=&
\sum_{n=0}^{N-1} \cos\left(\frac{2\pi}{N} kn\right)\left(a_{n+1}(t)+a_{n-1}(t)-2a_n(t)\right)
\nonumber \\
&=& -2 \left[ 1 - \cos \frac{2\pi}{N} k\right] \wht{a}(t,k)
\label{Z3} \\
{\cal C}\left(\demi\left(b_{n+1}(t)-b_{n-1}(t)\right)\right)(k) &=&
\demi \sum_{n=0}^{N-1} \cos\left(\frac{2\pi}{N} kn\right)\left(b_{n+1}(t)-b_{n-1}(t)\right) \nonumber \\
&=& \sin \left(\frac{2\pi}{N} k\right) \wht{b}(t,k)
\label{Z4}
\EEA
and also  the sine-transformations of the odd functions
\BEA
{\cal S}\left(b_{n+1}(t)+b_{n-1}(t)-2b_n(t)\right)(k) &=&
\sum_{n=0}^{N-1} \sin\left(\frac{2\pi}{N} kn\right)\left(b_{n+1}(t)+b_{n-1}(t)-2b_n(t)\right)
\nonumber \\
&=& -2 \left[ 1 - \cos \frac{2\pi}{N} k\right] \wht{b}(t,k)
\label{Z5} \\
{\cal S}\left(\demi\left(a_{n+1}(t)-a_{n-1}(t)\right)\right)(k) &=&
\demi \sum_{n=0}^{N-1} \sin\left(\frac{2\pi}{N} kn\right)\left(a_{n+1}(t)-a_{n-1}(t)\right)
\nonumber \\
&=& - \sin \left(\frac{2\pi}{N} k\right) \wht{a}(t,k)
\label{Z6}
\EEA

\noindent
{\bf Acknowledgements:}  MH is grateful to KIAS S\'eoul for warm hospitality, where a large part of this work was done.
We thank  N. Allegra, L. Berthier, J.-Y. Fortin, H. Park, U.C. T\"auber and M. Zannetti for useful discussions and/or correspondence.
This work was also partly supported by the Coll\`ege Doctoral franco-allemand Nancy-Leipzig-Coventry
({\it `Syst\`emes complexes \`a l'\'equilibre et hors \'equilibre'}) of UFA-DFH
and also by the National Research
Foundation of Korea (NRF) grant funded by the Korea
government (MSIP) (No. 2016R1A2B2013972).

%%%%%%%%%%%%%%%%%%%%%%%%%%%%%%%%%%%%%%%%%%%%%%%%%%%%%%%%%%%%%%%%%%%%%%%%%%%%%%%%%%
%\newpage


{\small

\begin{thebibliography}{999}

\bibitem{Abra65} M. Abramowitz and I.A. Stegun, {\it Handbook of Mathematical Functions}, Dover (New York 1965)

%\bibitem{Alme13} R.A.L. Almeida, S.O. Ferreira, T.J. Oliveira, F.D.A. Aar\~{a}o Reis,
%Phys. Rev. {\bf B89}, 045309 (2014) {\tt [arxiv:1312.1478]}.

%\bibitem{Alve14} S.G. Alvez, T.J. Oliveira and S.C. Ferreira, Phys. Rev. {\bf E90}, 020103(R) (2014).

%\bibitem{Anni06} A. Annibale and P. Sollich, J. Phys. {\bf A39}, 2853 (2006) {\tt [cond-mat/0510731]} ;\\
%                 A. Annibale and P. Sollich, J. Stat. Mech. P02064 (2009) {\tt [arXiv:0811.3168]}.

%\bibitem{Assi14} T.A. de Assis and F.D.A. Ar\~ao Reis, Phys. Rev. {\bf E89}, 062405 (2014) {\tt [arXiv:1401.6246]}.

\bibitem{Atis14} S. Atis, S. Saha, H. Auradou, S. Salin, L. Talon, Phys. Rev. Lett. {\bf 110}, 148301 (2013) {\tt [arXiv:1210.3518]}; \\
                 S. Atis, A.K. Dubey, D. Salin, L. Talon, P. Le Doussal, K.J. Wiese, Phys. Rev. Lett. {\bf 114}, 234502 (2015) {\tt [arxiv:1410.1097]}.

\bibitem{Bara95} A.L. Barab\'asi and H.E. Stanley, {\it Fractal concepts in surface growth}, Cambridge University Press (1995).

%\bibitem{Baum05} F. Baumann, M. Henkel, M. Pleimling and J. Richert, J. Phys. {\bf A38}, 6623 (2005) {\tt [cond-mat/0504243]}.

%\bibitem{Baum06} F. Baumann, S. Stoimenov and M. Henkel, J. Phys. {\bf A39}, 4095 (2006) {\tt [cond-mat/0510778]}.

%\bibitem{Baum06b} F. Baumann and M. Pleimling, J. Phys. {\bf A39}, 1981 (2006) {\tt [cond-mat/0509064]}.

%\bibitem{Baum07} F. Baumann, S.B. Dutta and M. Henkel, J. Phys. {\bf A40}, 7389 (2007) {\tt [cond-mat/0703445]}.

%\bibitem{Baum07b} F. Baumann, and M. Henkel, J. Stat. Mech. P01012 (2007) {\tt [cond-mat/0611652]}.

\bibitem{benN12} E. ben-Naim and P.L. Krapivsky, J. Phys. {\bf A45}, 455003 (2012) {\tt [arxiv:1209.0043]}.

\bibitem{Bray92} A.J. Bray and K. Humayun, Phys. Rev. Lett. {\bf 68}, 1559 (1992).

\bibitem{Bray94} A.J. Bray, Adv. Phys. {\bf 43}, 357 (1994). 

%\bibitem{Bray00} A.J. Bray, in M.E. Cates and M.R. Evans (eds), {\it Soft and fragile matter}, IOP Press (Bristol 2000); pp. 205-236.

%\bibitem{Bray13} A.J. Bray, S.N. Majumdar and G. Schehr, Adv. Phys. {\bf 62}, 225 (2013) {\tt [arXiv:1304.1195]}.

\bibitem{Berl52} T.H. Berlin and M. Kac, Phys. Rev. {\bf 86}, 821 (1952).

\bibitem{Bert97} L. Bertini and G. Giacomin, Comm. Math. Phys. {\bf 183}, 571 (1997).

\bibitem{Bert00} L. Berthier, Eur. Phys. J. {\bf B17}, 689 (2000) {\tt [arxiv:cond-mat/0003122]}.

%\bibitem{Bort11} D. Borthwick and S. Garibaldi, Notices Am. Math. Soc. {\bf 58}, 1055 (2011) {\tt [arXiv:1012.5407]}.

\bibitem{Burg74} J.M. Burgers, {\it The nonlinear diffusion equation: asymptotic solutions and statistical problems}, Reidel (Dordrecht 1974).

\bibitem{Bust07} S. Bustingorry, J. Stat. Mech. P10002 (2007) {\tt [arXiv:0708.2615]}.

%\bibitem{Cala05} P. Calabrese and A. Gambassi, J. Phys. {\bf A38}, R133 (2005) {\tt [cond-mat/0410357]}.

\bibitem{Cala11} P. Calabrese and P. Le Doussal, Phys. Rev. Lett. {\bf 106}, 250603 (2011) {\tt [arXiv:1104.1993]}.

\bibitem{Cala14} P. Calabrese, M. Kormos and P. Le Doussal, Europhys. Lett. {\bf 107}, 10011 (2014) {\tt [arXiv:1405.2582]}.

%\bibitem{Cane11} L. Canet, H. Chat\'e. B. Delamotte, N. Wschebor, Phys. Rev. {\bf E84}, 061128 (2011) {\tt [arxiv:1107.2289]}.

%\bibitem{Cann01} S.A. Cannas, D.A. Stariolo and F.A. Tamarit, Physica {\bf A294}, 362 (2001) {\tt [cond-mat/0010319]}.

%\bibitem{Cham06} C. Chamon, L.F. Cugliandolo and H. Yoshino, J. Stat. Mech. P01006 (2006) {\tt [cond-mat/0506297]}.

%\bibitem{Cham11} C. Chamon, F. Corberi and L.F. Cugliandolo, J. Stat. Mech. P08015 (2011) {\tt [arXiv:1105.2949]}.

%\bibitem{Chou09} Y.-L. Chou, M. Pleimling and R.K.P. Zia, Phys. Rev. {\bf E80}, 061602 (2009) {\tt [arXiv:0912.0062]}; \\
%Y.-L. Chou and M. Pleimling, J. Stat. Mech. P08007 (2010) {\tt [arXiv:1007.2380]} ;\\
%Y.-L. Chou and M. Pleimling, Physica {\bf A391}, 3585 (2012) {\tt [arXiv:1112.5867]}.

\bibitem{Coni94} A. Coniglio and M. Zannetti, Europhys. Lett. {\bf 10}, 575 (1989);\\
                 A. Coniglio, P. Ruggiero and M. Zanetti, Phys. Rev. {\bf E50}, 1046 (1994).

\bibitem{Coni90} A. Coniglio and M. Zannetti, Physica {\bf A163}, 325 (1990); \\
                 C. Amitrano, A. Coniglio, P. Meakin, M. Zannetti, Fractals {\bf 1}, 840 (1993).

\bibitem{Corl96} R.M. Corless, G.H. Gonnet, D.E.G. Hare, D.J. Jeffrey, D.E. Knuth, Adv. Compt. Math. {\bf 5}, 329 (1996).

\bibitem{Corw12} I. Corwin, Rand. Matrices Theory Appl. {\bf 1}, 1130001 (2012) {\tt [arXiv:1106.1596]}.

%\bibitem{Cour74} R. Courant and F. John, {\it Introduction to calculus and analysis, vol. 2}, Wiley (New York 1974)

%\bibitem{Cris03} A. Crisanti and F. Ritort, J. Phys. {\bf A36}, R181 (2003) {\tt [cond-mat/0212490]}.

%\bibitem{Cuei99} S. Cueille and C. Sire, Eur. Phys. J. {\bf B7}, 111 (1999) {\tt [cond-mat/9803014]}.

\bibitem{Cugl94} L.F. Cugliandolo and J. Kurchan, J. Phys. {\bf A27}, 5749 (1994) {\tt [cond-mat/9311016]}; \\
                 L.F. Cugliandolo, J. Kurchan and G. Parisi, J. Physique {\bf I4}, 1641 (1994) {\tt [cond-mat/9406053]}.

\bibitem{Cugl95} L.F. Cugliandolo and D. Dean, J. Phys. {\bf A28}, 4213 (1995) {\tt [cond-mat/9502075]}.

\bibitem{Cugl03} L.F. Cugliandolo, in J.-L. Barrat, M. Feiglman, J. Kurchan, J. Dalibard (eds),
{\it Slow relaxations and non-equilibrium dynamics in condensed matter}, Les Houches LXXVII,
Springer (Heidelberg 2003), pp. 367-521 {\tt [cond-mat/0210312]}.

%\bibitem{Cugl09} L.F. Cugliandolo, Physica {\bf A389}, 4360 (2009) {\tt [arXiv:0911.0771]}.

%\bibitem{Cugl11} L.F. Cugliandolo, J. Phys. {\bf A44}, 483001 (2011) {\tt [arXiv:1104.4901]}.

\bibitem{Daqu11} G.L. Daquila and U.C. T\"auber, Phys. Rev. {\bf E83}, 051107 (2011) {\tt [arXiv:1102.2824]}.

%\bibitem{Dohe94} J.P. Doherty, M.A. Moore, J.M. Kim, A.J. Bray, Phys. Rev. Lett. {\bf 72}, 2041 (1994).

%\bibitem{Dura09} X. Durang and M. Henkel, J. Phys. {\bf A42}, 395004 (2009) {\tt [arXiv:0905.4876]}.

\bibitem{Dutt08} S.B. Dutta, J. Phys. {\bf A41}, 395002 (2008) {\tt [arXiv:0806.3642]}.

%\bibitem{Ebbi08} M. Ebbinghaus, H. Grandclaude, M. Henkel, Eur. Phys. J. {\bf B63}, 81 (2008) {\tt [arxiv:0709.3220]}.

\bibitem{Edwa82} S.F. Edwards and D.R. Wilkinson, Proc. Roy. Soc. {\bf A381}, 17 (1982).

\bibitem{Fami85} F. Family and T. Vicsek, J. Phys. {\bf A18}, L75 (1985).

\bibitem{Fell71} W. Feller, {\it An introduction to probability theory and its applications}, vol. 2 (2$^{\rm nd}$ ed),
                 Wiley (New York 1971).

\bibitem{Fort12} J.-Y. Fortin and S. Mantelli, J. Phys. {\bf A45}, 475001 (2012) {\tt [arxiv:1208.2114]}.

\bibitem{Fusc02} N. Fusco and M. Zannetti, Phys. Rev. {\bf E66}, 066113 (2002) {\tt [cond-mat/0210502]}.

\bibitem{Fyod15} Y.V. Fyodorov, A. Perret and G. Schehr, J. Stat. Mech. P11017 (2015) {\tt [arxiv:1507.08520]}.

%\bibitem{Glau63} R.J. Glauber, J. Math. Phys. {\bf 4}, 294 (1963).

\bibitem{Godr00b} C. Godr\`eche and J.-M. Luck, J. Phys. {\bf A33}, 9141 (2000) {\tt [cond-mat/0001264]}.

%\bibitem{Godr02} C. Godr\`eche and J.-M. Luck, J. Phys. Cond. Matt. {\bf 14}, 1589 (2002) {\tt [cond-mat/0109212]}.

\bibitem{Godr13} C. Godr\`eche and J.-M. Luck, J. Stat. Mech. P05006 (2013) {\tt [arXiv:1302.4658]}.

\bibitem{Gwa92} L.-H. Gwa and H. Spohn, Phys. Rev. {\bf A46}, 844 (1992).

\bibitem{Halp95} T. Halpin-Healy and Y.-C. Zhang, Phys. Rep. {\bf 254}, 215 (1995).

\bibitem{Halp12} T. Halpin-Healy, Phys. Rev. Lett. {\bf 109}, 170602 (2012).

\bibitem{Halp13} T. Halpin-Healy, Phys. Rev. {\bf E88}, 042118 (2013); erratum {\bf E88}, 069903(E) (2013).

\bibitem{Halp14b} T. Halpin-Healy and Y. Lin, Phys. Rev. {\bf E89}, 010103(R) (2014) {\tt [arxiv:1310.8013]}.

\bibitem{Halp14} T. Halpin-Healy and G. Palansantzas, Europhys. Lett. {\bf 105}, 50001 (2014) {\tt [arxiv:1403.7509]}.

%\bibitem{Hase06} M.O. Hase and S.R. Salinas, J. Phys. {\bf A39}, 4875 (2006) {\tt [cond-mat/0512286]}.

%\bibitem{Hase12} M.O. Hase and M.J. de Oliveira, J. Phys. {\bf A45} 165003 (2012) {\tt [arXiv:1112.4893]}.

%\bibitem{Henk05} M. Henkel, A. Picone and M. Pleimling, Europhys. Lett. {\bf 68}, 191 (2005) {\tt [cond-mat/0404464]}.

%\bibitem{Henk06} M. Henkel, T. Enss and M. Pleimling, J. Phys. {\bf A39}, L589 (2006) {\tt [cond-mat/0605211]}.

%\bibitem{Henk07} M. Henkel and F. Baumann, J. Stat. Mech. P07015 (2007) {\tt [cond-mat/0703226]}.

%\bibitem{Henk09b} M. Henkel and M. Pleimling, J. Stat. Mech. P12012 (2009) {\tt [arXiv:0907.1642]}.

%\bibitem{Henk09} M. Henkel, H. Hinrichsen and S. L\"ubeck,
%{\it ``Non-equilibrium phase transitions vol. 1: absorbing phase transitions''}, Springer (Heidelberg 2009).

\bibitem{Henk10} M. Henkel and M. Pleimling, {\it ``Non-equilibrium phase transitions vol. 2:
                 ageing and dynamical scaling far from equilibrium''}, Springer (Heidelberg 2010).

\bibitem{Henk12} M. Henkel, J.D. Noh and M. Pleimling, Phys. Rev. {\bf E85}, 030102(R) (2012) {\tt [arxiv:1109.5022]}.

%\bibitem{Henk13} M. Henkel, Nucl. Phys. {\bf B869}, 282 (2013) {\tt [arXiv:1009.4139v2]}.

\bibitem{Henk15} M. Henkel and X. Durang, J. Stat. Mech. P05022 (2015) {\tt [arxiv:1501.07745]}.

%\bibitem{Houc02} B. Houchmandzadeh, Phys. Rev. {\bf E66}, 052902 (2002).

%\bibitem{Howa97} M. Howard and U.C. T\"auber, J. Phys. {\bf A30}, 7721 (1997) {\tt [cond-mat/9701069]}.

\bibitem{Huer12} M.A.C. Huergo, M.A. Pasquale, A.E. Bolz\'an, A.J. Arvia and P.H. Gonz\'alez, Phys. Rev. {\bf E82}, 031903 (2010); \\
                 M.A.C. Huergo, M.A. Pasquale, P.H. Gonz\'alez, A.E. Bolz\'an and A.J. Arvia, Phys. Rev. {\bf E84}, 021917 (2011); \\
                 M.A.C. Huergo, M.A. Pasquale, P.H. Gonz\'alez, A.E. Bolz\'an and A.J. Arvia, Phys. Rev. {\bf E85}, 011918 (2012).

\bibitem{Huer14} M.A.C. Huergo, N.E. Muzzio, M.A. Pasquale, P.H. Gonz\'alez, A.E. Bolz\'an and A.J. Arvia, Phys. Rev. {\bf E90}, 022706 (2014).

\bibitem{Igua09} J.L. Iguain, S. Bustingorry, A.B. Kolton, L.F. Cugliandolo, Phys. Rev. {\bf B80}, 094201 (2009) {\tt [arXiv:0903.4878]}; \\
                 S. Bustingorry, L.F. Cugliandolo and J.L. Iguain, J. Stat. Mech. P09008 (2007) {\tt [arxiv:0705.3348]}.

\bibitem{Imam12} T. Imamura and T. Sasamoto, Phys. Rev. Lett. {\bf 108}, 190603 (2014) {\tt [arXiv:1111.4634]}; \\
J. Stat. Phys. {\bf 150}, 908 (2013) {\tt [arXiv:1210.4278]}.

%\bibitem{Jans89} H.K. Janssen, B. Schaub and B. Schmittmann, Z. Phys. {\bf B73}, 539 (1989); \\
%H.K. Janssen, in G. Gy\"orgi {it et al.} (eds), {\it From phase transitions to chaos}, World Scientific (Singapour 1992).

\bibitem{Joyc72} G.S. Joyce, in C. Domb and M.S. Green (eds) {\it Phase transitions and critical phenomena}, Vol. 2,
Academic Press (London 1972), pp. 375ff.

\bibitem{Kall99} H. Kallabis and J. Krug, Europhys. Lett. {\bf 45}, 20 (1999) {\tt [cond-mat/9809241]}.

\bibitem{Kard86} M. Kardar, G. Parisi and Y.-C. Zhang, Phys. Rev. Lett. {\bf 56}, 889 (1986).

\bibitem{Kell16} J. Kelling, G. \'Odor and S. Gemming, J. Phys. {\bf A50}, 12LT01 (2017) {\tt [arxiv:1605.02620]}.

\bibitem{Kell17} J. Kelling, G. \'Odor and S. Gemming, {\tt [arxiv:1701.03638]}. 
\bibitem{Kell17b} J. Kelling, G. \'Odor and S. Gemming, Comp. Phys. Comm. {\bf 220}, 205 (2017) {\tt [arXiv:1705.01022]}.

\bibitem{Kenn06} R. Kenna, D.C. Johnston, W. Janke, Phys. Rev. Lett.{\bf 96}, 115701 (2006)  {\tt [arxiv:cond-mat/0605162]}; \\
                 R. Kenna, D.C. Johnston, W. Janke, Phys. Rev. Lett. {\bf  97} 155702 (2006) {\tt [arxiv:cond-mat/0608127]};
                                                    erratum Phys. Rev. Lett. {\bf 97} (2006) 169901 (2006); \\
                 R. Kenna, in Yu. Holovatch (ed) {\it Order, Disorder and Criticality: Advanced Problems of Phase Transition Theory}, vol. 3,
                 World Scientific (Singapour 2013), p. 1 {\tt [arXiv:1205.4252]}.

%\bibitem{Kim89} J.M. Kim and J.M. Kosterlitz, Phys. Rev. Lett. {\bf 62}, 2289 (1989).

%\bibitem{Kim14} S.-W. Kim and J.M. Kim, J. Stat. Mech. P07005 (2014).

%\bibitem{Klos12} T. Kloss, L. Canet, N. Wschebor, Phys. Rev. {\bf E86}. 051124 (2012) {\tt [arxiv:1209.4650]}.

\bibitem{Krec97} M. Krech, Phys. Rev. {\bf E55}, 668 (1997) {\tt [cond-mat/9609230]}; erratum {\bf E56}, 1285 (1997).

\bibitem{Krie10} T. Kriecherbauer and J. Krug, J. Phys. {\bf A43}, 403001 (2010) {\tt [arXiv:0803.2796]}.

\bibitem{Krug97} J. Krug, Adv. Phys. {\bf 46}, 139 (1997).

\bibitem{Kurc02} J. Kurchan, Phys. Rev. {\bf E66}, 017101 (2002) {\tt [arxiv:cond-mat/0110628]}.

\bibitem{Lambert1758} J.H. Lambert, Acta Helv. {\bf 3}, 128 (1758); \\
                 L. Euler, Acta Acad. Scient. Petropol. {\bf II}, 29 (1779) [paper E532, printed 1783] 
                           and {\it Opera Omnia, Series Prima}, vol. 6, Teubner (Leipzig 1921);\\
                 G. Polya und G. Szeg\"o, {\it Aufgaben und Lehrs\"atze aus der Analysis}, 2 B\"ande, 4. Auflage, Springer (Heidelberg 1970/71);\\ 
                 G. Polya and G. Szeg\"o, {\it Problems and theorems in analysis} 2 vols., 5$^{\rm th}$ ed., Springer (New York 1998).                 

\bibitem{Ligg85} T. Liggett, {\it Interacting particle systems}, Springer (Heidelberg 1985).

%\bibitem{Leuz09} L. Leuzzi, J. Non-Crystalline Solids {\bf 355}, 686 (2009) {\tt [arXiv:0810.1405]}.

\bibitem{Lewi52} H.W. Lewis and G.H. Wannier, Phys. Rev. {\bf 88}, 682 (1952); erratum {\bf 90}, 1131 (1953).

%\bibitem{Luck85} J.-M. Luck, Phys. Rev. {\bf B31}, 3069 (1985).

%\bibitem{Maju96} S.N. Majumdar, A.J. Bray, S.J. Cornell and C. Sire, Phys. Rev. Lett. {\bf 77}, 3704 (1996) {\tt [cond-mat/9606123]}.

\bibitem{Mall15} K. Mallick, Physica {\bf A418}, 17 (2015) {\tt [arxiv:1412.6258]}.

%\bibitem{Maun97} J. Maunuksela, M. Myllys, O.-P. K\"ahk\"onen, J. Timonen, N. Provatas,
%M. J. Alava, and T. Ala-Nissila, Phys. Rev. Lett. {\bf 79}, 1515 (1997); \\
%M. Myllys, J. Maunuksela, M. Alava, T. Ala-Nissila, J. Merikosi and J. Timonen, Phys. Rev. {\bf E64}, 036101 (2001) {\tt [cond-mat/0105234]} .

%\bibitem{Marc08} U. Marconi, B. Marini, A. Puglisi, Phys. Rep. {\bf 461}, 111 (2008) {\tt [arXiv:0803.0719]}.

\bibitem{Maze06} G.F. Mazenko, {\it Non-equilibrium statistical mechanics}, Wiley (New York 2006).

\bibitem{Mens12} A. Menshutin, Phys. Rev. Lett. {\bf 108}, 015501 (2012).

%\bibitem{Mini12} D. Minic, C. Vaman and C. Wu, Phys. Rev. Lett. {\bf 109}, 131601 (2012) {\tt [arxiv:1207.0243]}.

\bibitem{Moha09} F. Mohammadi, A.A. Saberi, S. Rouhani, J. Phys. Cond. Matt. {\bf 21}, 375110 (2009) {\tt [arxiv:0905.0820]}.

\bibitem{Odor14} G. \'Odor, J. Kelling, S. Gemming, Phys. Rev. {\bf E89}, 032146 (2014) {\tt [arxiv:1312.6029]}.

%\bibitem{Onsa44} L. Onsager, Phys. Rev. {\bf 65}, 117 (1944).

\bibitem{Oono88} Y. Oono and S. Puri, Mod. Phys. Lett. {\bf B2}, 861 (1988).

%\bibitem{Paes03} M. Paessens and M. Henkel, J. Phys. {\bf A36}, 8993 (2003) {\tt [cond-mat/0306171]}.

%\bibitem{Paes04} M. Paessens and G.M. Sch\"utz, J. Phys. {\bf A37}, 4709 (2004) {\tt [cond-mat/0311568]}.

\bibitem{Pagn15} A. Pagnani, G. Parisi, Phys. Rev. {\bf E92}, 010101(R) (2015)  {\tt [arxiv:1611.08445]}. 

\bibitem{Pico02} A. Picone and M. Henkel, J. Phys. {\bf A35}, 5575 (2002) {\tt [cond-mat/0203411]}.

%\bibitem{Pico04} A. Picone and M. Henkel, Nucl. Phys. {\bf B688}, 217 (2004) {\tt [cond-mat/0402196]}.

%\bibitem{Popk14} V. Popkov, J. Schmidt and G.M. Sch\"utz, Phys. Rev. Lett. {\bf 112}, 200602 (2014) {\tt [arxiv:1312.5920]}

%\bibitem{Popk15a} V. Popkov, J. Schmidt and G.M. Sch\"utz, J. Stat. Phys. {\bf 160}, 835 (2015) {\tt [arxiv:1410.8026]}

%\bibitem{Popk15b} V. Popkov, A. Schadschneider, J. Schmidt and G.M. Sch\"utz,
%Proc. Nat. Acad. Sci. {\bf 112}, 12645 (2015) {\tt [arxiv:1505.04461]}.

\bibitem{Prae00} M. Pr\"ahofer, H. Spohn, Physica {\bf A279}, 342 (2000)  {\tt [arXiv:cond-mat/9910273]}. 

%\bibitem{Prudnikov2} A.P. Prudnikov, Yu.A. Brychkov, O.I. Marichev,
%{\it ``Integrals and series vol 2: special functions''}, Gordon and Breach (New York 1986).

%\bibitem{Rodr15} E.A. Rodrigues, B.A. Mello, F.A. Oliveira, J. Phys. {\bf A48}, 035001 (2015).

\bibitem{Ronc78} G. Ronca, J. Chem. Phys. {\bf 68}, 3737 (1978).

\bibitem{Roet06} A. R\"othlein, F. Baumann and M. Pleimling, Phys. Rev. {\bf E74}, 061604 (2006) {\tt [cond-mat/0609707]};
erratum {\bf E76}, 019901(E) (2007).

\bibitem{Sasa10} T. Sasamoto and H. Spohn, Phys. Rev. Lett. {\bf 104}, 230602 (2010) {\tt [arXiv:1002.1883]}.

%\bibitem{Salv96} R.C. Salvarezza, L V\'azquez, H. M\'{\i}guez, R. Mayoral, C. L\'opez, F. Meseguer, Phys. Rev. Lett. {\bf 77}, 4572 (1996);\\
%L. V\'azquez, R.C. Salvarezza, A.J. Arvia, Phys. Rev. Lett. {\bf 79}, 709 (1997).

%\bibitem{Schi99} P.L. Schilardi, O. Azzaroni, R.C. Salvarezza and A.J. Arvia, Phys. Rev. {\bf B59}, 4638 (1999).

\bibitem{Singh85} S. Singh, R.K. Pathriah, Phys. Rev. {\bf B31}, 4483 (1985).

%\bibitem{Sire04} C. Sire, Phys. Rev. Lett. {\bf 93}, 130602 (2004) {\tt [cond-mat/0406333]}.

%\bibitem{Snep92} K. Sneppen, Phys. Rev. Lett. {\bf 92}, 3539 (1992).

\bibitem{Somf03} E. Somfai, R.C. Ball, N.E. Bowler, L.M. Sander, Physica {\bf A325}, 19 (2003) {\tt [arxiv:cond-mat/0210637]}.

%\bibitem{Spoh15} H. Spohn and G. Stoltz, J. Stat. Phys. {\bf 160}, 861 (2015) {\tt [arxiv:1410.7896]}.

\bibitem{Stru78} L.C.E. Struik, {\it Physical ageing in amorphous polymers and other materials}, Elsevier (Amsterdam 1978).

\bibitem{Take11} K.A. Takeuchi, M. Sano, T. Sasamoto and H. Spohn,
Sci. Reports {\bf 1}:34 (2011) {\tt [arxiv:1108.2118]}; \\
K.A. Takeuchi and M. Sano, Phys. Rev. Lett. {\bf 104}, 230601 (2010) {\tt [arxiv:1001.5121]}.

\bibitem{Take12} K.A. Takeuchi and M. Sano, J. Stat. Phys. {\bf 147}, 853 (2012) {\tt [arxiv:1203.2530]}.

\bibitem{Take14} K.A. Takeuchi, J. Stat. Mech. P01006 (2014) {\tt [arxiv:1310.0220]}.

\bibitem{Take17} K.A. Takeuchi,  {\tt [arxiv:1708.06060]}.

%\bibitem{Tang92} L.-H. Tang and H. Leschhorn, Phys. Rev. {\bf A45}, R8309 (1992); \\
%S.V. Buldyrev, A.-L. Barab\'asi, F. Caserta, S. Havlin, H.E. Stanley and T. Vicsek, Phys. Rev. {\bf A45}, R8313 (1992).

\bibitem{Taeu14} U.C. T\"auber, {\it Critical dynamics: a field-theory approach to equilibrium and non-equilibrium scaling behavior},
Cambridge University Press (Cambridge 2014).

\bibitem{Taeu17} U.C. T\"auber, Ann. Rev. Cond. Matter Phys. {\bf 8}, 1 (2017) {\tt [arxiv:1604.04487]}. 

\bibitem{Vinc07} E. Vincent, in M. Henkel, M. Pleimling, R. Sanctuary (eds) {\it Ageing and the glass transition}, Springer Lecture Notes in Physics
{\bf 716}, Springer (Heidelberg 2007), p. 1, {\tt [arxiv:cond-mat/0603583]}.

\bibitem{Wald17} S. Wald, G.T. Landi, M. Henkel, J. Stat. Mech. (at press) {\tt [arxiv:1707.06273]}.

%\bibitem{Warn94} S.O. Warnaar, P.A. Pearce, K.A. Seaton and B. Nienhuis, J. Stat. Phys. {\bf 74}, 469 (1994) {\tt [hep-th/9305134]}.

%\bibitem{Wies98} K.J. Wiese, J. Stat. Phys. {\bf 93}, 143 (1998) {\tt [cond-mat/9802068]}.

\bibitem{Wio13} H. Wio, R.R. Deza, C. Escudero, J.A. Revelli, Papers in Phys. {\bf 5}, 050010 (2013) {\tt [arxiv:1401.6425]}.

\bibitem{Wio17} H. Wio, M.A. Rodriguez, R. Gallego, J.A. Revelli, A. Al\'es, R.R. Deza, Frontiers in Physics {\bf 4}, 52 (2017). 

%\bibitem{Yeun96} C. Yeung, M. Rao and R.C. Desai, Phys. Rev. {\bf E53}, 3073 (1996) {\tt [cond-mat/9409108]}.

\bibitem{Yunk13} P.J. Yunker, M.A. Lohr, T. Still, A. Borodin, D.J. Durian, A.G. Yodh,
Phys. Rev. Lett. {\bf 110}, 035501 (2013) {\tt [arxiv:1209.4137]}; \\
Comment: M. Nicoli, R. Cuerno, M. Castro, Phys. Rev. Lett. {\bf 111}, 209601 (2013);
P.J. Yunker {\it et al.}, Phys. Rev. Lett. {\bf 111}, 209602 (2013).

%\bibitem{Zamo89} A.B. Zamolodchikov, Adv. Stud. Pure Math. {\bf 19}, 641 (1989).

%\bibitem{Zipp00} W. Zippold, R. K\"uhn and H. Horner, Eur. Phys. J. {\bf B13}, 531 (2000) {\tt [cond-mat/9904329]}.


\end{thebibliography}

}
\end{document}